\documentclass[preprint,aps,showpacs,preprintnumbers,amsmath,amssymb,nofootinbib]
{revtex4}
\usepackage{epstopdf}
\usepackage{epsfig}
\usepackage{graphicx}
\usepackage{color}
\begin{document}
\begin{flushright}
\end{flushright}
\newcommand{\be}{\begin{equation}}
\newcommand{\ee}{\end{equation}}
\newcommand{\bea}{\begin{eqnarray}}
\newcommand{\eea}{\end{eqnarray}}
\newcommand{\nn}{\nonumber}
\def\CP{{\it CP}~}
\def\cp{{\it CP}}
\title{\large Impact of  lepton flavour universality violation on CP violation sensitivity of long baseline neutrino oscillation experiments}
\author{Soumya C. and R. Mohanta }
\affiliation{School of Physics, University of Hyderabad, Hyderabad - 500 046, India }
\begin{abstract}
  The observation of neutrino oscillation as well as the recent  experimental result on lepton flavor universality (LFU) violation  in  $B$ meson decays 
are indications of new physics  beyond the  Standard Model. Many theoretical models, which are introduced in the literature as an extension of SM 
to explain these observed deviations in LFU, lead to new kind of interactions so-called non-standard interaction (NSI) between the elementary particles.   
In this paper, we consider a model with an additional $Z'$ boson (which is quite 
 successful  in explaining the  
observed LFU anomalies) and  analyze its effect in the lepton flavour violating (LFV)  
 $B_d\to \tau^\pm e^\mp$ decay modes. From the present upper bound  of the
$B_d\to \tau^\pm e^\mp$ branching ratio, we obtain the constraints on the new physics 
parameters, which are related to the corresponding NSI parameters in the neutrino sector by $SU(2)_L$ symmetry.  
These new parameters are expected to have potential implications in the neutrino oscillation studies and in this work  
we investigate the possibility of observing  the effects of these interactions in the currently running and upcoming long-baseline
experiments, i.e., NO$\nu$A and DUNE respectively.
\end{abstract}
\maketitle
\section{Introduction}
The Standard Model of particle physics, which seems to provide a complete picture of interaction and dynamics of elementary particles 
with the discovery of Higgs boson at LHC \cite{lhc1}, predicts the equality of electroweak couplings of electron and muons so-called Lepton Flavor Universality (LFU). 
However, the observation of neutrino oscillation, which allows mixing between different lepton families of neutrinos, 
implies that family lepton number is violated, and the violation in LFU are indications of new physics (NP) beyond the SM. Moreover, 
the  deviations in recent observation of the violation of LFU in semileptonic $B$ decays, both in the case of $b \to c$ charged-current  as
well as in the case of $b \to s$ neutral current transitions, also point towards physics beyond the SM.  
These results can be summarized as follows:\\
$\bullet$  About $4.0\sigma$ deviation of $\tau/l$  universality ($l = \mu, e)$ in $b \to c$ transitions \cite{RD}, i.e., 
\bea
&&R ({D^*}) = \frac{{\rm Br} (B \to D^* \tau \nu_\tau)}{{\rm Br} (B \to D^* l \nu_l)}=0.316 \pm 0.016 \pm 0.010\;,\nn\\
&&R({D}) = \frac{{\rm Br} (B \to D \tau \nu_\tau)}{{\rm Br} (B \to D
 l \nu_l)}=0.397 \pm 0.040 \pm 0.028 \;,
\eea 
from their corresponding SM values $R(D^*)|_{\rm SM}=0.252 \pm 0.003$ \cite{fajfer} and $R(D)|_{\rm SM}=0.300 \pm 0.008$ \cite{RD1}.
Since these
decays are mediated at tree level in the SM, relatively
large new physics  contributions are necessary  to
explain these deviations.

$\bullet $ Observation of $2.6\sigma $ deviation of $\mu/e$ universality in the 
dilepton invariant mass bin $1~{\rm GeV}^2 \leqslant q^2 \leqslant 6~ {\rm GeV}^2$
  in
$b \to s$ transitions \cite{RK}:
\bea
R_K = \frac{{\rm Br}(B \to K \mu^+ \mu^-)}
{{\rm Br}(B \to K e^+ e^-)}
= 0.745_{-0.074}^{+0.090} \pm 0.036 ,
\eea
from the SM prediction $R_K^{\rm SM}=1.0003 \pm 0.0001$.

$\bullet $ CMS recently also searched for the decay $h \to \tau \mu$
and found a non-zero result of ${\rm Br} (h \to \tau \mu) = 0.84^{+0.39}_{-0.37}$ \cite{cms-1}
which disagrees by about $2.4 \sigma$ from 0, i.e. from the SM
value.\\ 
These deviations from the SM have triggered a series of theoretical speculations about
possible existence of NP beyond the SM. Some of the prominent  NP models which
can explain these deviations from the SM 
are: models with an extra $Z'$ boson \cite{z-prime} and/or additional
Higgs doublets \cite{thdm}, models with leptoquarks \cite{lepto} etc. The  observation of lepton flavour non-universality 
effects also provide the possibility of the observation of lepton flavour violating (LFV) decays \cite{glashow}.
Although so far, there is no concrete evidence of LFV decays but there exist  strict upper bounds in many
LFV decays such as $\mu \to e \gamma $ $\mu \to eee$, etc \cite{pdg}. Various dedicated experiments are already planned to 
search for LFV decays.
In this paper, we would like to see the implications of the LFV interactions in various long-baseline neutrino oscillation experiments.
In other words, we would like to explore whether it is possible to observe 
these effects in the long-baseline neutrino oscillation experiments or not. 
In particular, we will focus on the NP contributions which could affect only to the $\tau$ sector.
This is particularly interesting as the tauonic $B$ decays provide an excellent probe of new physics
because of the involvement of heavy $\tau$ lepton. There are a few deviations observed in the leptonic/semileptonic $B$ decays with a $\tau$
in the final state.
We consider the model with an additional $Z'$ boson, which can mediate flavour changing neutral current (FCNC) transitions
at tree level.
$Z'$ gauge bosons, which are associated with as extra $U(1)'$ gauge symmetry,
are predicted theoretically in
many extensions of the SM \cite{GUT},  such as grand unified theories (GUTs), 
left-right symmetric models, $E_6$ model,
supersymmetric models,
 superstring theories etc.
Although the $U(1)'$ charges are in general family-universal but it is not mandatory to be so, 
and the family non-universal $Z'$  has been introduced in
some models, such as in  $E_6$ model \cite{e6}.
On the experiment side also there are  many efforts undergoing to search for the $Z'$ 
directly at the LEP, Tevatron,
and LHC. With the assumption that the coupling of $Z'$  to the SM fermions 
are similar to those of the SM $Z$ boson,
the direct searches for the $Z'$ can be performed in the dilepton events. 
At this stage, the lower mass limit has
been set as 2.9 TeV at the 95\% C.L. with  8 TeV data set by using $e^+ e^-$ and 
$\mu^+ \mu^-$  \cite{Atlas-1} events and this value becomes 1.9 TeV using the $\tau^+ \tau^-$ events
\cite{Atlas-2}. However, such constraints from the
LHC would not be valid if the $Z'$ boson couples very weakly with the leptons, and thus one has 
to rely on the hadronic channels. 

The paper is organized as follows. In section II, we discuss the possible hints of new physics  from $B$ meson decays and  extract the constraints on the  lepton flavor violating new NP parameters in the charged lepton sector  from the from the decay mode  $B_d \to \tau^\pm e^\mp$. 
These parameters are in general related to the corresponding 
NP parameters in the neutrino sector by the $SU(2)_L$
gauge symmetry.  The basic formalism of neutrino oscillation including NSI effects are briefly discussed in section III.
 In section  IV, we study the effect of NSI parameters on $\nu_{e}$ appearance oscillation 
probability and the search for the new CP violating signals at long-baseline experiments is presented
in section V. Section VI contains the summary and conclusions.
 
\section{New physics effects from $B$ meson decays} 
In this section, we would like to see the possible interplay of new physics  in the $\tau$-lepton sector considering the decay channels of $B$
meson. For this purpose, we first consider the leptonic decay channel $ B^- \to \tau^- \bar \nu$.
During the last few years, there has been a systematic disagreement between the
experimental and SM predicted value for the branching ratio of $B \to \tau \nu$ mode.
The branching ratio for $ B ^- \to \tau \nu_\tau$ is given as
\bea
{\rm Br}(B^- \to \tau \bar \nu_\tau) = \frac{G_F^2 }{8 \pi} |V_{ub}|^2 \tau_{B^-} f_B^2 m_B  m_\tau^2 \left (1- \frac{m_\tau^2}{m_B^2} \right )^2.
\label{br}
\eea
This mode is very clean and the only non-perturbative quantity involved in the expression for branching ratio (\ref{br}) is the decay constant of $B$ meson. 
However, there is still a tension
between the  exclusive and inclusive value of $V_{ub}$ at the level of $3 \sigma$. 
This mode has been precisely measured \cite{pdg} with a value 
\bea
{\rm Br}(B^- \to \tau^- \bar \nu_\tau) = (1.14 \pm 0.27) \times 10^{-4}\;.\label{expt}
\eea 
The latest result from Belle Collaboration \cite{belle-1}
\bea
  {\rm Br}(B^- \to \tau^- \bar \nu_\tau)  = (1.25 \pm 0.28 \pm 0.27) \times  10^{-4}\;,
\eea
also in the line of  the previous measurements. 
Since there is an uncertainty between the $|V_{ub}|$ values extracted from exclusive and inclusive modes, we use
the SM fitted value of its branching ratio from UTfit collaboration \cite{UTfit}
\bea
{\rm Br}(B^- \to \tau^- \bar \nu_\tau) = (0.84 \pm 0.07) \times 10^{-4}\;.\label{sm}
\eea 
This value agrees well with the experimental value (\ref{expt}). However, the central values of these two results differ significantly.
One can eliminate the $V_{ub}$ dependence
completely by introducing the LFU probing ratio 
\bea
 R_{\tau/l}^\pi = \frac{\tau_{ B^0 }}{\tau_{B^- }} \frac{{\rm Br}(B^- \to
\tau^- \bar \nu_\tau)}{{\rm Br}(B^0 \to \pi^0 l^- \bar \nu_l)} = 0.73 \pm 0.15\;,
\eea 
which has around $2.6\sigma$  deviation from its  SM prediction
of $R_{\tau/l}^{\pi,SM} = 0.31(6)$ \cite{fajfer}. Thus,  these deviations
may be considered as the smoking gun signal of  new physics associated with the tauonic sector. We then proceed to obtain the bound on the lepton 
flavor violating new physics parameter associated with the $\tau$ lepton from   the decay mode $B_d \to \tau^\pm e^\mp$.

\subsection{ Extraction of the NP parameter from the lepton flavour violating  decay process $B_d \to \tau^\pm e^\mp$   } \label{etbound}
The violation of lepton flavour universality in principle can induce lepton flavour violation. In this section, we will consider the lepton
flavour violating  decay process $B_d \to \tau^\pm e^\mp$,   which is induced by flavour changing neutral current interactions. 
As an example, here we will consider a simple 
and well-motivated model,
which would induce lepton flavour violating interactions at the tree level, is the model with an additional $Z'$ boson. 
Many SM extensions  often involve the  presence of an extra $U(1)'$ gauge symmetry and the corresponding gauge boson is 
generally known as the $Z'$ boson. Here we consider the model which can
induce the lepton flavour violating decays both in the down quark sector and the charged lepton sector \cite{z-prime, damir} at the tree level.    
Thus, in this   model the coupling of $Z'$ boson  to down type quarks and  charged leptons can be written generically as
\bea
{\cal L}\supset g' \Big [\eta_{db}^{L} \bar d \gamma^\mu P_L b  
+\eta_{d b }^{R} \bar d \gamma^\mu P_R b+\eta_{e \tau }^{L} \bar e \gamma^\mu P_L \tau  
+\eta_{e \tau }^{R} \bar e \gamma^\mu P_R \tau   \Big ]\;,
\eea
where $g'$ is the new $U(1)'$ gauge coupling constant, $\eta_{db }^{L/R}$  are the vector/axial vector FCNC couplings of 
$\bar d  b$ quark-antiquark pair to the $Z'$ boson and  $\eta_{e \tau }^{L,R}$ are the LFV  parameters.
\begin{figure}[!htb]
\begin{center}
\includegraphics[scale=0.8]{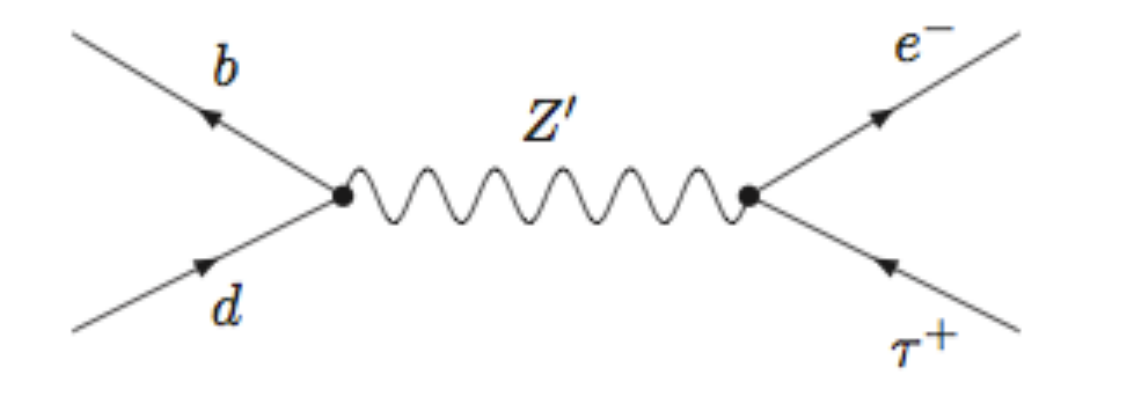}
\caption{Feynman diagram for $B_d \to e^- \tau^+ $ in the model
with  $Z'$ boson, where the blobs represent the tree
level FCNC couplings  of $Z'$ boson.}\label{feyn}
\end{center}
\end{figure}

The constraint on the  LFV coupling $\eta_{e \tau}$ can be obtained from the lepton flavour violating
$B$ decay mode $B_d \to \tau^\pm e^\mp$. In the SM this decay mode is  loop-suppressed
with tiny neutrino mass in the loop. However, in the $Z'$ model  it can occur at tree level,
described by the quark level transition $b \to d \tau^\pm e^\mp$
and is expected to have  significantly large branching ratio.
The  Feynman
diagram for this process in the $Z'$ model is shown in Fig. 1, where the blobs
represent the tree level FCNC coupling of $Z'$ boson.
The present upper limit on its branching ratio  is $2.8 \times 10^{-5}$.
The effective Hamiltonian describing  this process in the $Z'$ model can be given as
\bea
{\cal H}_{eff} = \frac{G_F}{\sqrt 2} 
 \left ( \frac{g' M_Z}{gM_{Z'}}\right )^2[\bar d
\gamma^\mu (\eta_{db }^L- \eta_{db }^R \gamma_5) b]
[\bar e
\gamma_\mu (\eta_{e\tau }^L- \eta_{e\tau }^R \gamma_5) \tau]\;,\label{hm-Bd}
\eea 
 where $M_{Z'}$ is the mass of $Z'$ boson.
In order to evaluate the transition
amplitude we use the following matrix element 
\bea \langle 0 
| \bar d \gamma^\mu(1-\gamma_5) b |B_d \rangle = -i f_B p_B^\mu \;, \label{dc-Bd} \eea 
where $f_{B}$ is the decay constant of $B$
meson and $p_B$ its momentum.  Thus, with eqns. (\ref{hm-Bd}) and (\ref{dc-Bd}),  one can
obtain the transition amplitude for the process $B_d \to \tau^- e^+$ as 
\bea {\cal M}(B_d \to \tau^- e^+)=-\frac{G_F}{\sqrt 2}\left ( \frac{g' M_Z}{gM_{Z'}}\right )^2  i f_B \eta_{db }^R~ p_B^\mu
 [\bar e \gamma_\mu   (\eta_{e\tau }^L- \eta_{e\tau }^R \gamma_5) \tau]\;,
\eea 
and the corresponding branching ratio is given as 
\bea
{\rm Br}(B_d \to \tau^\pm e^\mp)= \frac{G_F^2 \tau_B}{16 \pi}\left ( \frac{g' M_Z}{gM_{Z'}}\right )^4 |\eta_{db }^R|^2 
(|\eta_{e\tau }^L|^2+ |\eta_{e\tau }^R|^2)
f_B^2 m_\tau^2 m_B
\left (1- \frac{m_\tau^2}{m_B^2} \right )^2,\label{taue}
\eea
where $\tau_B$ is the lifetime of $B$ meson. In order to find out the bound on the LFV couplings $\eta_{e \tau}^{L,R}$, we need to know
the value of the parameter $\eta_{db}$, which can be obtained from the decay process $B_d \to \mu^+ \mu^-$. The branching ratio for
this decay mode  has been recently
measured by the LHCb \cite{lhcb19} and CMS \cite{cms19} collaborations and the present world average value \cite{nature}  is given as
\bea
{\rm Br}(B_d \to \mu^+ \mu^-)
=\left (3.9_{-1.4}^{+1.6} \right ) \times 10^{-10}\;.\label{br-exp}
\eea
The corresponding SM value has been precisely calculated including the corrections of ${\cal O}(\alpha)$ and ${\cal O}(\alpha_s^2)$
with value  \cite{bsll} 
\bea
{\rm Br}(B_d \to \mu^+ \mu^-)|_{\rm SM}
&=&\left (1.06 \pm 0.09  \right ) \times 10^{-10}\;.\label{brmu}
\eea
Although the SM predicted value is in agreement with the experimental result but it does not exclude the possible existence of
new physics as the central values of these two results differ significantly. The effective Hamiltonian 
describing this process is given as
\bea
{\cal{H}}_{eff}
= -\frac{G_F }{\sqrt{2} } \frac{ \alpha }{2 \pi} V_{tb} V_{td}^*
C_{10} [ \bar d \gamma^\mu (1-\gamma_5)b] [\bar \mu \gamma_\mu \gamma_5 \mu ],\label{ham-mu}
\eea
where  $C_{10}$ is the Wilson coefficient and its value at the $m_b$ scale is given as   $C_{10}=-4.245$.
The corresponding Hamiltonian in the $Z'$ model is given as
\bea
{\cal H}_{eff}^{Z'} = \frac{G_F}{\sqrt 2} 
 \left ( \frac{g' M_Z}{gM_{Z'}}\right )^2[\bar d
\gamma^\mu (\eta_{db }^L- \eta_{db }^R \gamma_5) b]
[\bar \mu
\gamma_\mu (C_V^\mu- C_A^\mu \gamma_5) \mu]\;,\label{hm-Bd1}
 \eea
where $C_V^\mu$ and $C_A^\mu$ are the vector and axial-vector
couplings of the $Z'$ boson to $\mu^-\mu^+$ pair. 
Including the contribution arising from the $Z'$ exchange to the SM amplitude, one can write the  amplitude for $B_d \to \mu \mu$ process as 
\bea
{\cal M}(B_d \to \mu^+ \mu^-) 
&=&  i\frac{G_F}{ {\sqrt 2}}\frac{\alpha}{ \pi}i V_{tb}V_{td}^*  f_{B} m_{B} m_\mu C_{10}  [\bar \mu \gamma_5 \mu] \left (
1+   \frac{g'^2 M_Z^2}{g^2 M_{Z'}^2} \frac{2 \pi \eta_{db }^R C_A^\mu  }{\alpha V_{tb}V_{td}^* C_{10}}\right )\nn\\
&=& {\cal M}^{SM} \left (1+   \frac{g'^2 M_Z^2}{g^2 M_{Z'}^2} \frac{2 \pi \eta_{db }^R C_A^\mu  }{\alpha V_{tb}V_{td}^* C_{10}}\right )
\;.\label{eq-Bd}
\eea
Thus, from Eq. (\ref{eq-Bd}), one can obtain the  branching ratio as
\bea
{\rm Br}(B_d \to \mu \mu)= {\rm Br}(B_d \to \mu \mu)^{SM}
\left |1+   \frac{g'^2 M_Z^2}{g^2 M_{Z'}^2} \frac{2 \pi \eta_{db }^R C_A^\mu  }{\alpha V_{tb}V_{td}^* C_{10}}\right |^2.\label{eq32}
\eea
Assuming  the  axial-vector coupling of $Z'$ to muon pair, i.e., $C_A^\mu$ 
 has the same form as the the corresponding SM $Z$ boson coupling to
fermion-antifermion pair   with value
 $C_A^\mu=-1/2$. 
Now  with Eqn. (\ref{eq32}) and considering  1-$\sigma$ range of  experimental and SM predicted  branching ratios from (\ref{brmu}) and (\ref{br-exp}),  
the constraint on the parameter $\eta_{db }^R$ is found to be 
\bea
0.006 \leq |\eta_{db }^R| \leq 0.014,\label{eta}
\eea
for $M_{Z'}$=1 TeV,  where we have used the particle masses and CKM elements from \cite{pdg}. 
Using this allowed range of $|\eta_{db }^R|$, the bounds on the LFV couplings $\eta_{e\tau}^{L,R}$  
can be obtained by comparing (\ref{taue}) with the corresponding branching ratio ${\rm Br}(B_d \to \tau e) < 2.8 \times 10^{-5}$ \cite{pdg}
as 
\bea
|\eta_{e\tau}^{L}|=|\eta_{e\tau}^{R}| < 19.2\;,~~~~{\rm for}~~|\eta_{db }^R|=0.014\;,\label{epset}
\eea
where we have considered $\eta_{e\tau}^{L}=\eta_{e\tau}^{R}$.
These couplings can be redefined in terms of another set of  new couplings as $\varepsilon_{e \tau}=(g'^2M_Z^2/g^2 M_{Z'}^2) \eta_{e \tau} $,  which can
give the relative  NP strength in comparison to SM ones  as 
\bea
|\varepsilon_{e\tau}^{L}|=|\varepsilon_{e\tau}^{R}| < 0.16\;,~~~~{\rm for}~~|\eta_{db }^R|=0.014\;,\label{bnd}
\eea
for $g'\simeq g$ and a TeV scale $Z'$ boson, i.e., $M_{Z'} \simeq 1$ TeV.
Since these parameters are related to the corresponding
NSI parameters of the neutrino sector by the $SU(2)_L$ symmetry, we now proceed to see their implications in
various long baseline neutrino oscillation experiments. Analogously, one can obtain the bounds on the NSI couplings $\varepsilon_{e \mu}$ from $B_d \to e \mu $ decay, which are expected to be of the same order as 
$\varepsilon_{e \tau} $.
\section{Neutrino oscillation in presence of NSIs}
 Neutrino oscillation \cite{SK,SNO,KL,T2K,DC,DB,RENO} has been established as a leading mechanism behind the flavour transition of 
neutrinos, which provides strong evidence for neutrino mass and mixing. Moreover, the three flavor framework of neutrino oscillation is  
very successful in explaining observed  experimental results except few results at very short baseline experiments. Nevertheless, there are 
few parameters in  oscillation framework, which are still not known, for instance the neutrino mass ordering, CP violating phase and the octant 
of atmospheric mixing angle. The main objective of the currently running and future up-coming long-baseline experiments is to determine these unknowns. Though these  experiments will take  a long time to collect the whole oscillation data, phenomenological studies can make predictions on the sensitivity of these experiments, which ultimately help to extract improved oscillation data. In this context, some phenomenological studies regarding the sensitivity of long-baseline experiments can be found in our recent works \cite{CS1,CS2,DKN}. At this point of time, where the neutrino physics entered into precision era, it is crucial to understand the effect of sub-leading  contributions such as Non-standard interactions (NSIs) of neutrinos  on the sensitivities of long-baseline neutrino oscillation
 experiments. It is well-known that NSIs  of neutrinos \cite{NSI-1,NSI-2}, which  derived from various extensions of the SM,  can affect neutrino propagation, production, and detection mechanisms which are commonly known as propagation, source and detector NSIs. However, in this paper, we mainly focus on propagation NSIs and their effect on neutrino oscillation. The Lagrangian  corresponds to NSIs during the propagation of neutrino  is given by \cite{NSI-L}, 
\begin{equation}
{\cal L}_{\rm NSI} = -2\sqrt{2}G_F\varepsilon_{\alpha\beta}^{fC}(\overline{\nu}_\alpha \gamma^\mu P_L\nu_\beta)(\overline{f} \gamma_\mu P_C f) \,, 
\end{equation}
where $G_F$ is the Fermi coupling constant,  $\varepsilon_{\alpha\beta}^{fC}$ are the  new coupling constants known as NSI parameters, $f$ is  
fermion and $P_C=(1\pm\gamma_5)/2$ are the  right ($C = R$) and left ($C = L$)  chiral projection  operators. The NSI contributions which are 
relevant while neutrino propagate through the earth are those coming from the interaction of neutrinos with matter  ($e$, $u$ and $d$), since the 
earth matter is made up of these fermions only. Therefore, the effective NSI parameter is given by
\begin{equation}
\varepsilon_{\alpha\beta} = \sum_{f=e,u,d}\frac{n_f}{n_e} \varepsilon_{\alpha\beta}^{f} \,,\label{eps}
\end{equation}
where $\varepsilon^f_{\alpha\beta}= \varepsilon_{\alpha\beta}^{fL} + \varepsilon_{\alpha\beta} ^{fR}$, $n_f$ is the number density of the fermion $f$ and $n_e$ the 
number density of electrons in earth. For earth matter, we can assume that the number densities of electrons, protons and neutrons are equal, i.e, 
$n_n \approx n_p =n_e$.
Therefore, one can write $\varepsilon_{\alpha \beta}$ as \cite{epsiloneff-1}
\bea
\varepsilon_{\alpha \beta} \approx \sqrt{ \sum_{C} {(\varepsilon_{\alpha \beta}^{eC})^2 + (3\varepsilon_{\alpha \beta}^{uC})^2 + 
(3\varepsilon_{\alpha \beta}^{dC})^2}}\;.\label{bd}
\eea
 Thus, with Eqns. (\ref{bnd}) and  (\ref{bd}), the bound on 
the NSI parameter $\varepsilon_{e \tau}$ is found to be
\bea
\varepsilon_{e \tau} < 0.7\;,\label{eps-1}
\eea
where we have assumed that either left-handed or right-handed couplings  
would be present at a given time.

NSIs and their consequences can be studied in  both model-dependent and -independent approaches by which one can obtain the model-dependent and -independent bounds on the NSI parameters. 
Recently,  considering the model independent approach, we have studied   
 the effect of lepton flavor violating NSIs on physics potential of long-baseline experiments \cite{CS-nsi}. 
 Moreover, the recent works on the effect of NSI on the measurements of various neutrino oscillation experiments can be seen in 
\cite{AdG-nsi,PC-nsi,PH-nsi,1601.07730,1601.00927,MM-nsi,SKR-nsi,SKA-nsi,MG-nsi}. 
Since, we focus on model-dependent approach in this paper, we consider the  LVF decays of $B$ meson in $Z'$ model to get the bound on NSI parameter as discussed in Section \ref{etbound}. 
There are many works in the literature, which are dealt with extensive study of  model-dependent NSI parameters and 
their effect on neutrino oscillation experiments \cite{Md-YF,Md-DVF}. However, in this work
we focus on the lepton flavor violating NSI parameter, where the bound is obtained from the  LFV  decays of $B$ meson in a $Z'$ model and check its effect on the  measurements of CP 
violation at the long baseline experiments like  NO$\nu$A and DUNE. This would provide an indirect signal  for the existence of $Z'$ boson coming from the long-baseline neutrino experiment results.
\subsection{Basic formalism with NSIs}
The effective Hamiltonian describing the propagation of  neutrinos  through matter in  the standard three flavor framework  is given by
\begin{eqnarray}
H_{SO} &=&  H_{0} + H_{M} \nonumber\\
&=&\frac{1}{2E} U\cdot {\rm diag}(0,\Delta m^2_{21},\Delta m^2_{31}) \cdot U^{\dagger} +  {\rm diag}(V_{CC},0,0)\;,
\end{eqnarray}
where $H_{0}$ is the  Hamiltonian in vacuum, $\Delta m^2_{ji} = m^2_{j}-m^2_{i}$ is neutrino mass squared difference, $H_{M}$ 
is the Hamiltonian responsible for matter effect, $V_{CC}=\sqrt{2}G_{F}n_{e}$ is the matter potential and $U$ is the PMNS mixing matrix which is described 
by three mixing angles ($\theta_{12},\theta_{13},\theta_{23}$) and one CP violating phase ($\delta_{CP}$) is given by
    \begin{equation}
    U_{PMNS}= \left( \begin{array}{ccc} c^{}_{12} c^{}_{13} & s^{}_{12}
    c^{}_{13} & s^{}_{13} e^{-i\delta} \\ -s^{}_{12} c^{}_{23} -
   c^{}_{12} s^{}_{13} s^{}_{23} e^{i\delta} & c^{}_{12} c^{}_{23} -
   s^{}_{12} s^{}_{13} s^{}_{23} e^{i\delta} & c^{}_{13} s^{}_{23} \\
   s^{}_{12} s^{}_{23} - c^{}_{12} s^{}_{13} c^{}_{23} e^{i\delta} &
    -c^{}_{12} s^{}_{23} - s^{}_{12} s^{}_{13} c^{}_{23} e^{i\delta} &
   c^{}_{13} c^{}_{23} \end{array} \right),
    \label{standpara}
    \end{equation}
where $c_{ij} = \cos(\theta_{ij})$ and $s_{ij} = \sin(\theta_{ij})$.
The  NSI Hamiltonian, which describes the new  interactions between the matter particles as neutrinos propagate through matter  is given by
\begin{equation}
H_{NSI} = V_{CC}
 \begin{pmatrix}
  \varepsilon_{ee} & \varepsilon_{e\mu} & \varepsilon_{e\tau} \\
  \varepsilon^*_{e\mu} & \varepsilon_{\mu\mu} & \varepsilon_{\mu\tau} \\
  \varepsilon^*_{e\tau} & \varepsilon^*_{\mu\tau} & \varepsilon_{\tau\tau}
 \end{pmatrix},
 \label{nsi}
\end{equation}
where $\varepsilon_{\alpha\beta} =|\varepsilon_{\alpha\beta}|e^{i\delta_{\alpha \beta}}$ are the complex NSI parameters. Then the neutrino 
oscillation probability in presence of NSI is given by
\begin{equation}
P_{(\nu_{\alpha} \rightarrow \nu_{\beta})} = \left |\langle {\nu_{\beta}}e^{-i(H_{SO}+H_{NSI})L} \rangle {\nu_{\alpha}}\right |^2.
\end{equation}
In this paper, we focus on lepton flavor violating NSIs, i.e., the effects of the off-diagonal elements of the matrix (\ref{nsi}). Moreover, constraints from terrestrial experiments show that the muon sector is strongly constrained \cite{modelbound}, so that one can set $\varepsilon_{e \mu}$ and $\varepsilon_{\mu \tau}$ to zero. Therefore,
in our analysis we consider only the contributions from the NSI parameter  $\varepsilon_{e\tau}$ and use a conservative value
for $\varepsilon_{e\tau}$ as $\varepsilon_{e\tau}\approx 0.3$, consistent with the bound obtained from lepton flavour violating $B$ meson decays,
as shown in Eqn. (\ref{eps-1}). 
\section{Numerical analysis}
\subsection{Effect of NSI on oscillation probability and event spectra}
\begin{figure}[!htb]
\begin{center}
\includegraphics[width=8cm,height=8cm]{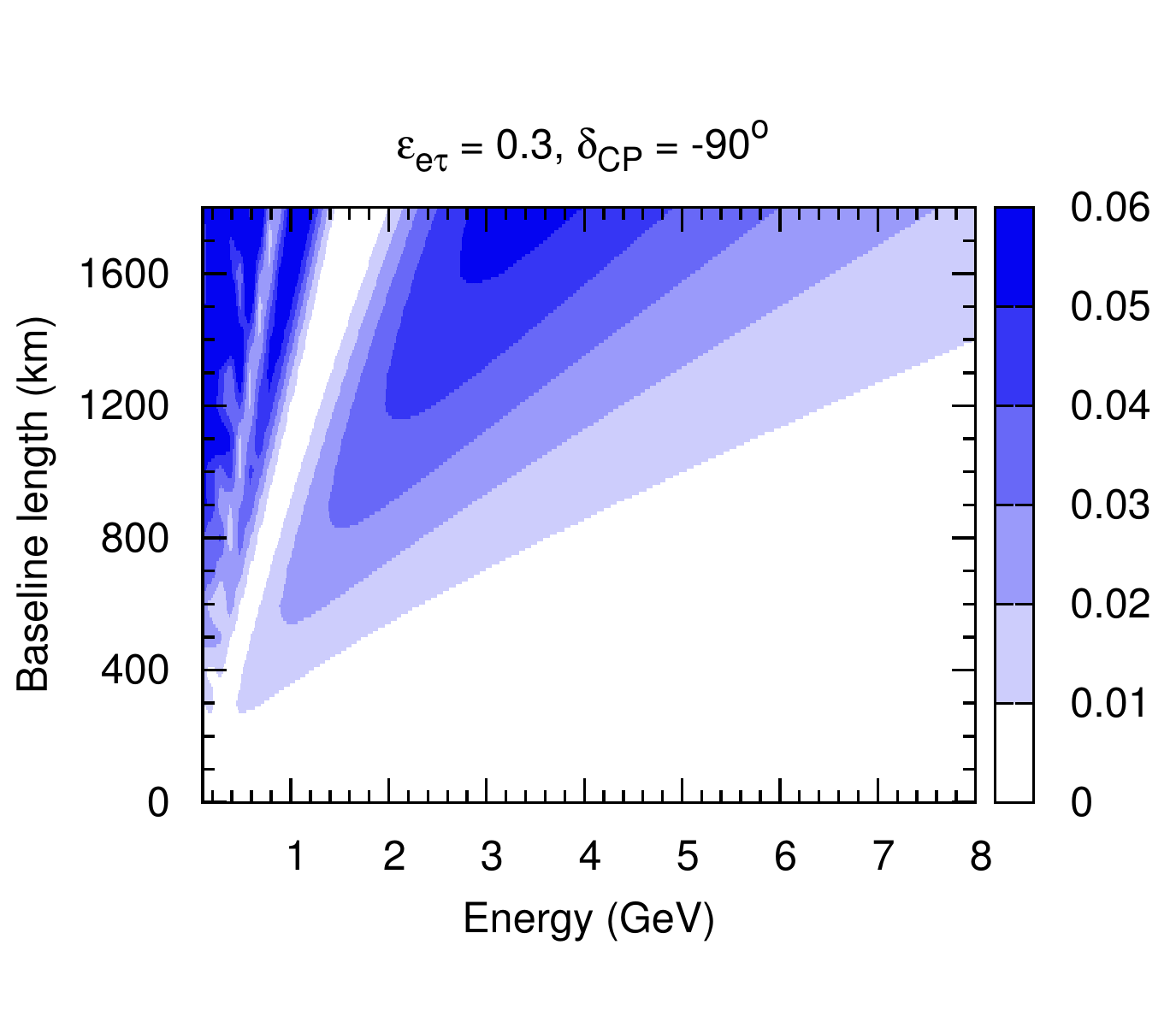}
\hspace*{0.2 true cm}
\includegraphics[width=8cm,height=8cm]{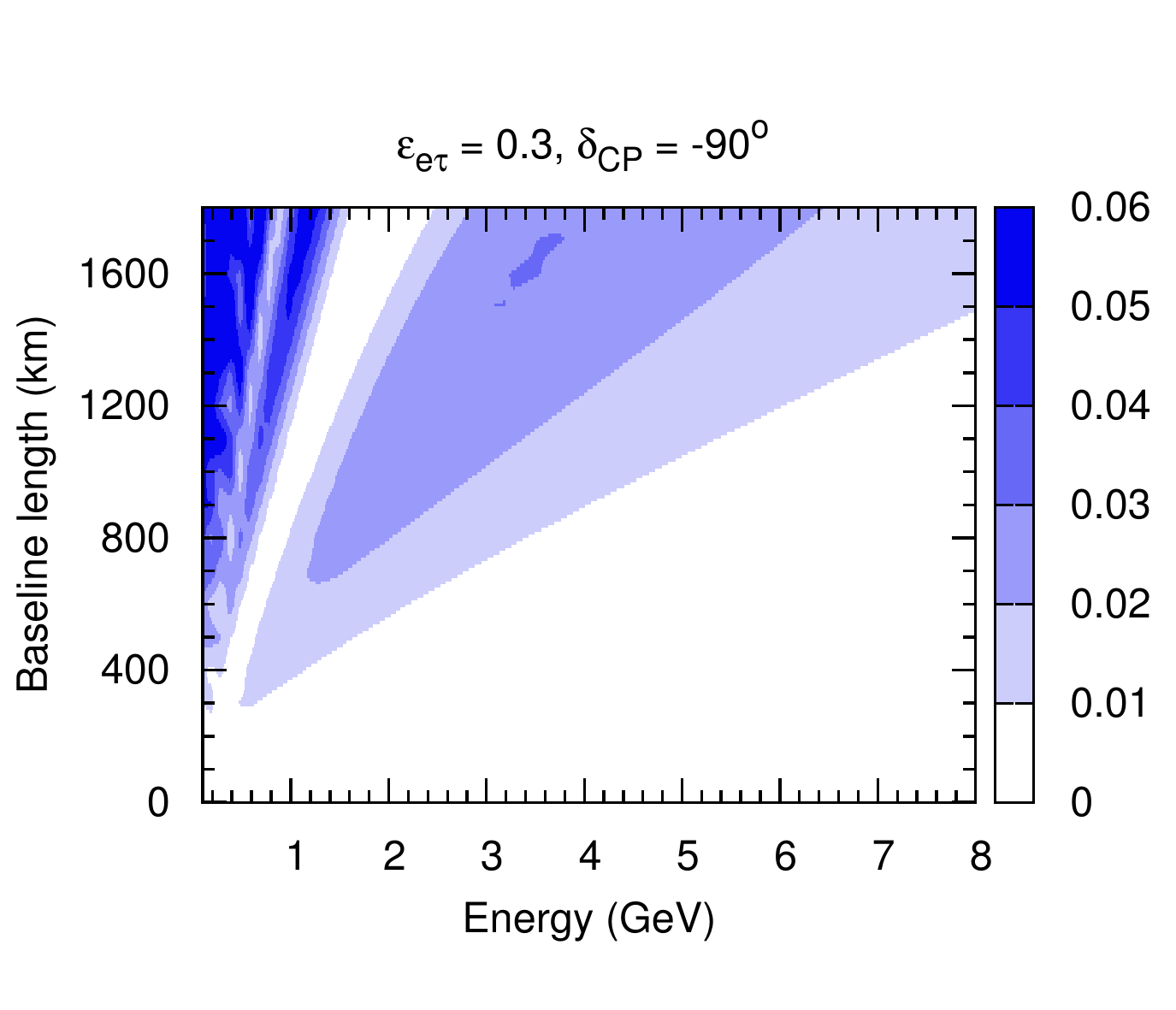}
\end{center}
\caption{ The $\Delta P = |P_{NSI}-P_{SI}|$ as a function of neutrino energy and baseline length. The left (right) panel corresponds to Normal (Inverted) hierarchy.}
\label{DP}
\end{figure}

In this section, we discuss the effect of NSI parameter on the neutrino oscillation probability as well as on the event spectra of long baseline experiments like NO$\nu$A and DUNE. 
We use GLoBES package \cite{Huber:2004-1,Huber:2009-2} for our  analysis. We also use snu plugin \cite{snu1,snu2} to incorporate Non-standard physics in GLoBES.
\begin{table}[h]
    \begin{center}
    \vspace*{0.1 true in} 
    \begin{tabular}{|l|c|c|}\hline
    ~Expt. setup& \text{NO$\nu$A} & \text{DUNE}\\
   & \cite{{nova,nova-1,sanjib}}& \cite{{lbne,lbne-2}}\\\hline
    Detector &~Scintillator~&~Liquid Argon~\\
    Beam Power(MW) & 0.77 & 0.7\\
    Fiducial mass(kt) & 14 & 40\\
    Baseline length(km) & 810 & 1300 \\
    Running time (yrs)&6 (3$\nu$+3$\bar{\nu}$)&10 (5$\nu$+5$\bar{\nu}$) \\
    \hline
    \end{tabular} 
    \caption{The experimental specifications.}
    \label{experiments}
    \end{center} 
    \end{table}  
The specifications of the long baseline experiment that we consider in this paper are given in the  Table \ref{experiments} and the true value of oscillation parameters that we use in our calculations are given in Table \ref{true-osc}.   
\begin{table}[h]
    \begin{center}
    \vspace*{0.1 true in}
    \begin{tabular}{|c|c|}\hline
    Oscillation Parameter & True Value\\ \hline
    $\sin^2\theta_{12}$ & 0.32 ~ \\\hline
    $\sin^2 2\theta_{13}$ & 0.1 ~ \\\hline
    $\sin^2 \theta_{23}$ & 0.5, 0.41 (LO), 0.59 (HO) ~ \\\hline
    $\Delta m_{atm}^2$ & $2.4 \times 10^{-3} ~{\rm eV}^2$ for NH ~ \\
    & $-2.4 \times 10^{-3} ~{\rm eV}^2$ for IH ~ \\\hline
    $\Delta m_{21}^2$ & $7.6 \times 10^{-5}~ {\rm eV}^2$ ~ \\\hline
    $\delta_{CP} $ & $-90^\circ$  ~ \\\hline
    \end{tabular}
    \caption{The true values of oscillation parameters considered in the simulations.}
    \label{true-osc}
    \end{center}
    \end{table} 
To show the  effect of NSI parameter $\varepsilon_{e\tau}$ on oscillation probability, we obtain $\Delta P = |P_{NSI}-P_{SI}|$ (where $P_{NSI(SI)}$ denotes the probability with Non-standard (Standard) interactions)  for different baseline length and energy using the neutrino oscillation parameters as given in Table \ref{true-osc}. The contour plots for $\Delta P$ as a function of neutrino energy and baseline length are given in the Fig.~\ref{DP}. The different shades in the figure correspond to different ranges of $\Delta P$. From the figure, we can see that  $\Delta P \in$ (0.02,0.03) and (0.04,0.05) for   NO$\nu$A ($L= 810$ km and $E = 2$ GeV) and DUNE ($L= 1300$ km and $E = 2.5$ GeV) respectively for NH, whereas for IH,  $\Delta P \in$ (0.02,0.03) for both NO$\nu$A and DUNE. This implies that the non-standard interactions can affect the measurement of oscillation parameters at NO$\nu$A and DUNE experiments significantly.   
   
\begin{figure}[!htb]
\begin{center}
\includegraphics[width=8cm,height=6cm]{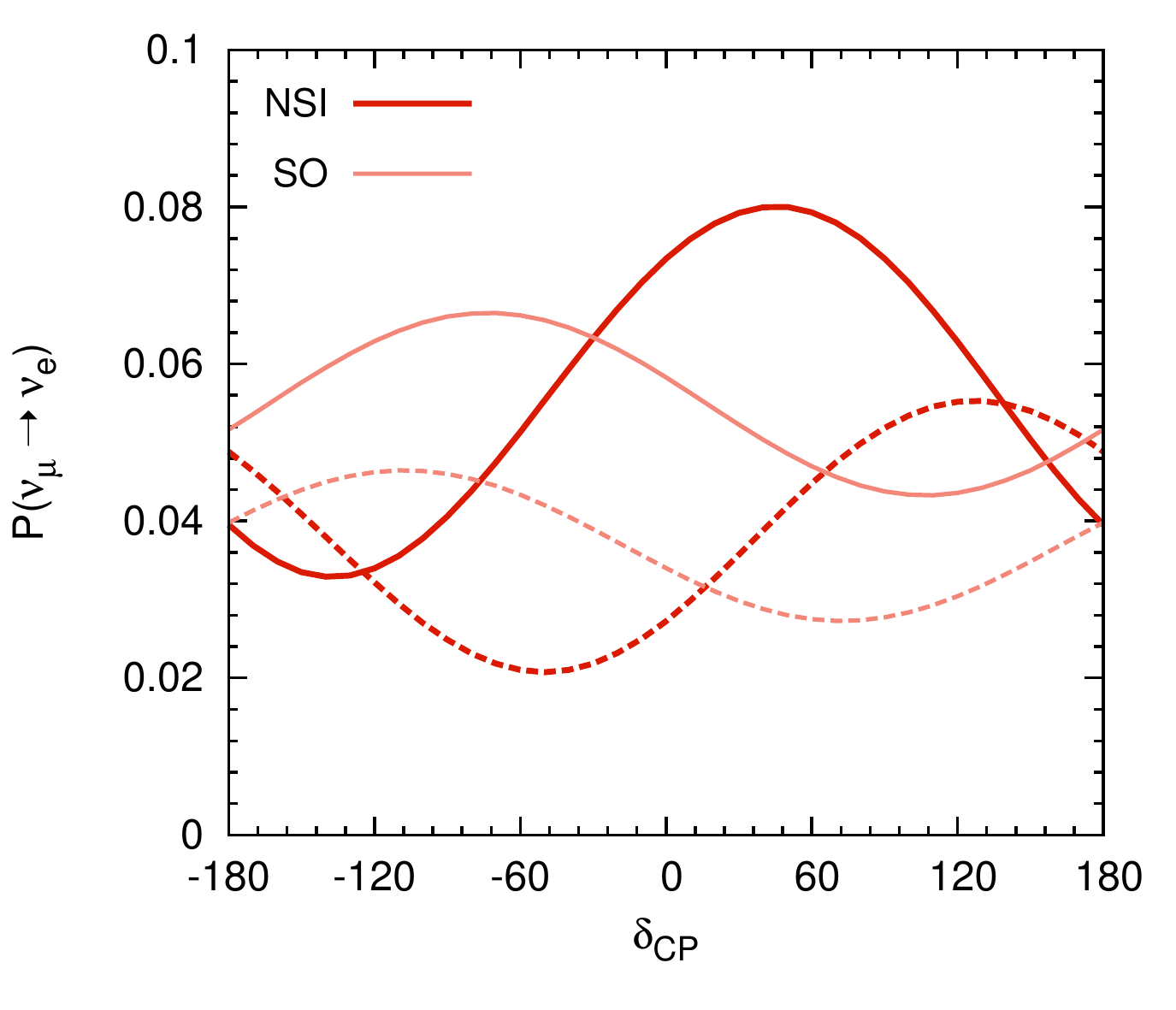}
\includegraphics[width=8cm,height=6cm]{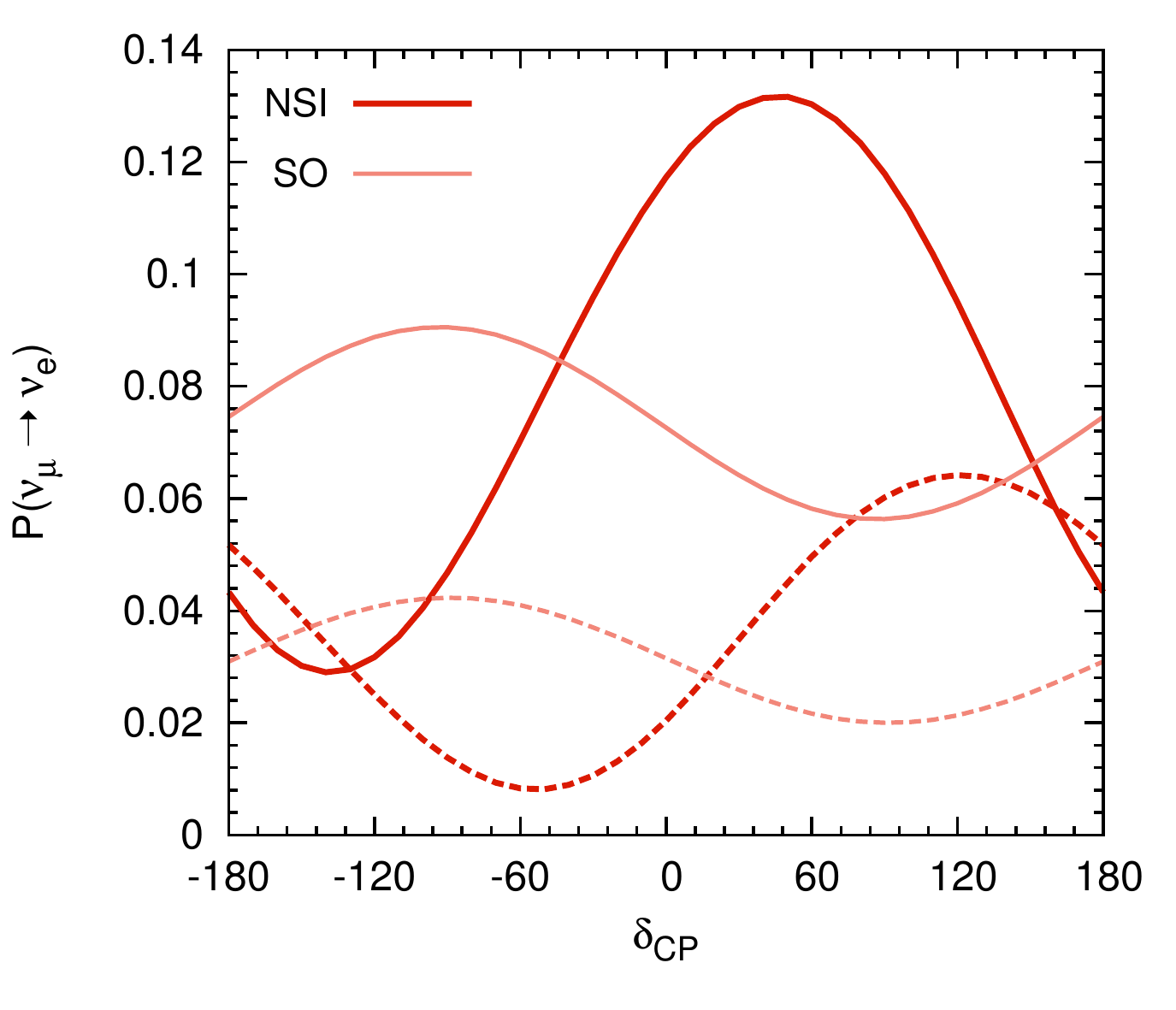}
\end{center}
\caption{The left (right) panel shows the appearance oscillation probability for  NO$\nu$A (DUNE). The dark (light) coloured curves represent the oscillation probability in the presence (absence) of NSI for  $\delta _{e\tau} = 0$. The solid (dashed) curves correspond to NH (IH). }
\label{app-disapp}
\end{figure}

\begin{figure}[!htb]
\begin{center}
\includegraphics[width=5.4cm,height=4.5cm]{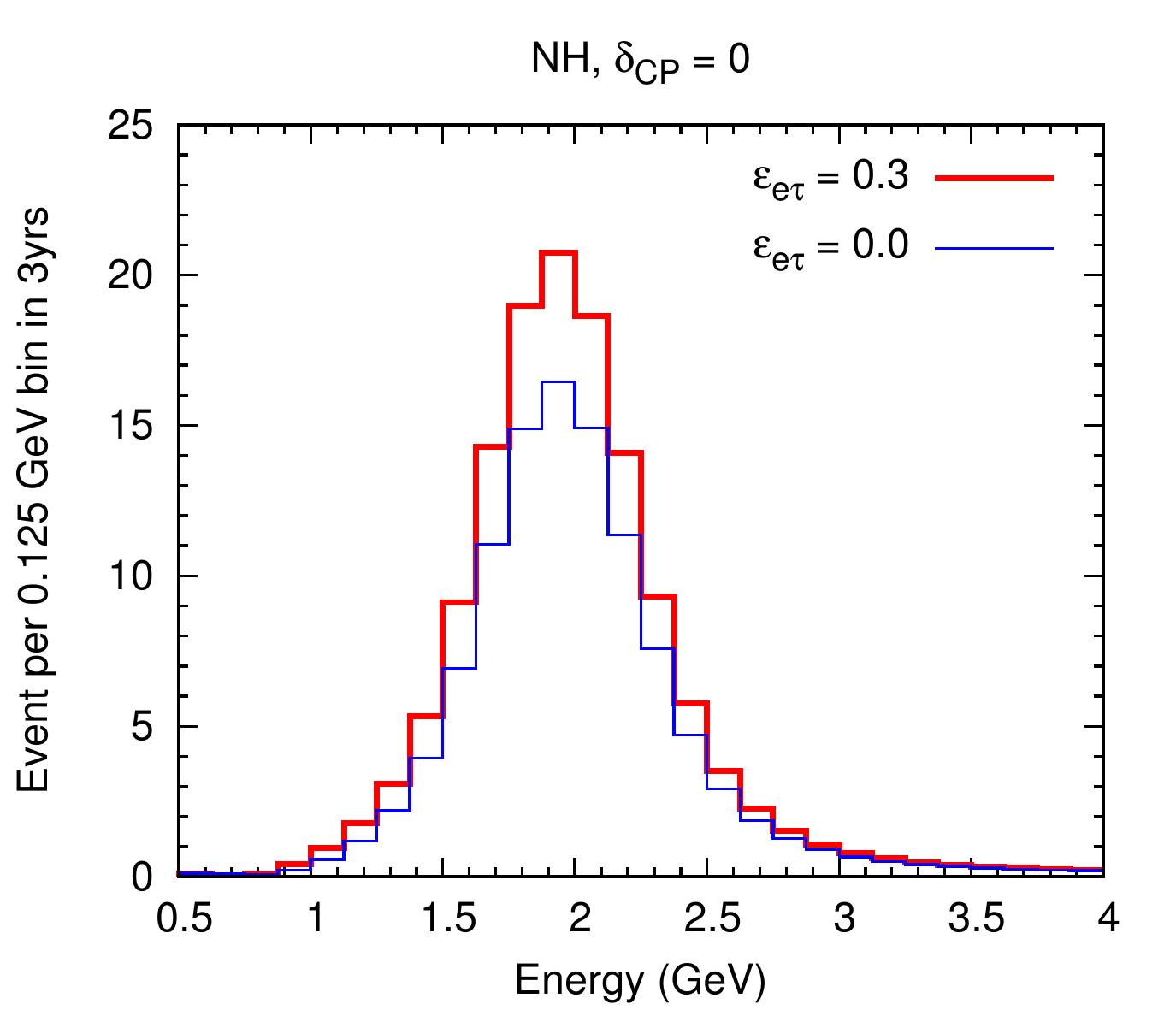}
\includegraphics[width=5.4cm,height=4.5cm]{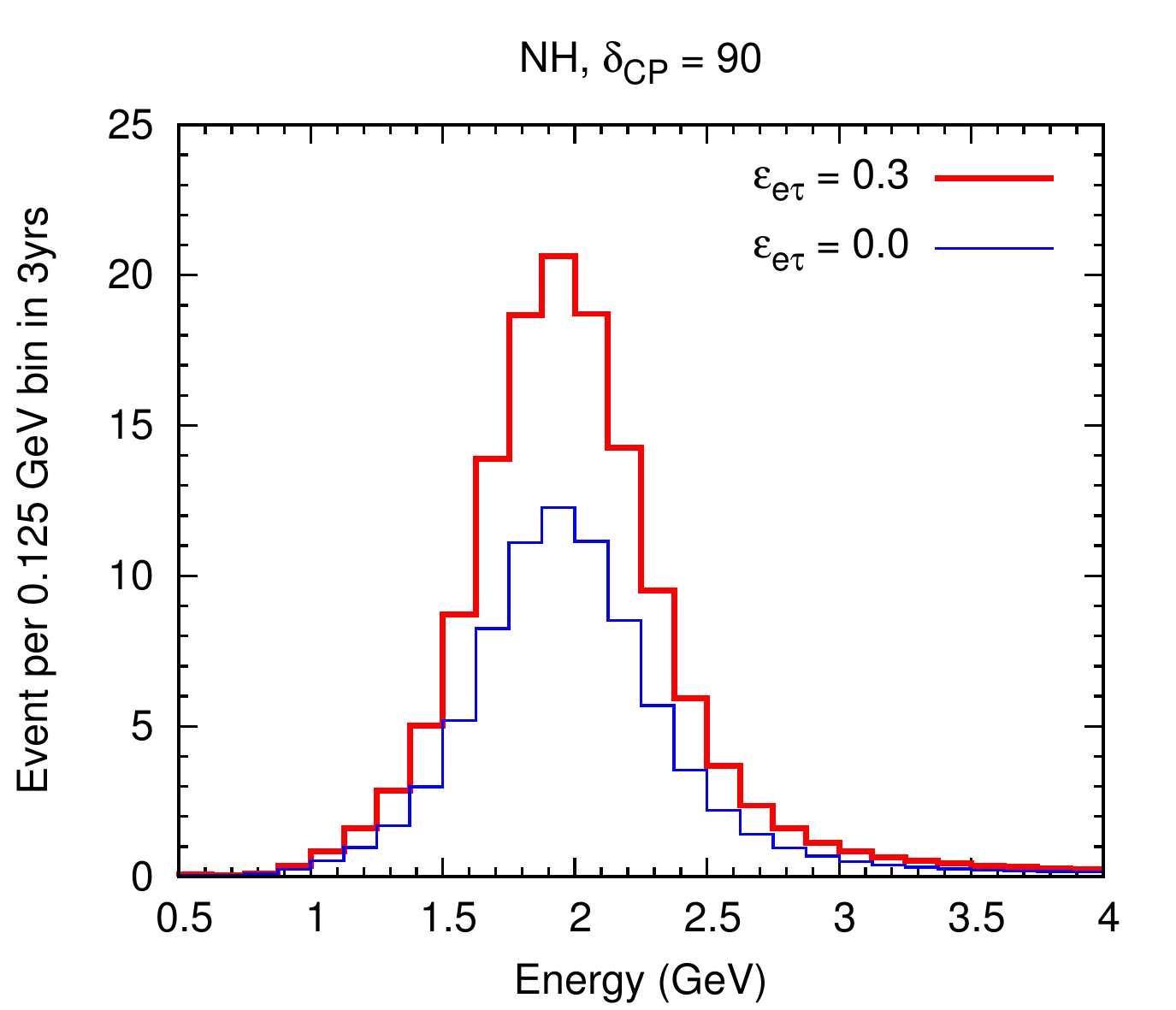}
\includegraphics[width=5.4cm,height=4.5cm]{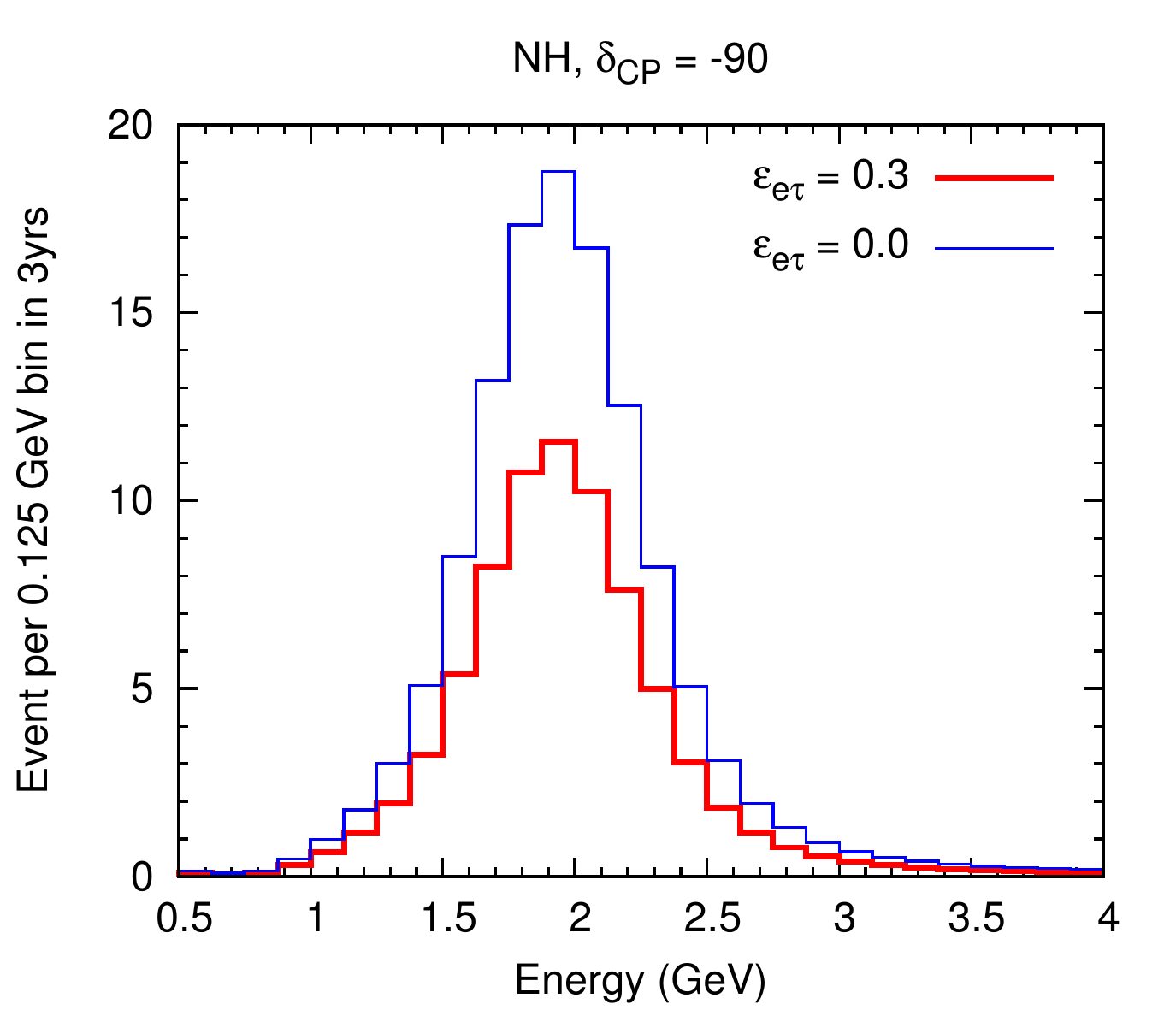}
\includegraphics[width=5.4cm,height=4.5cm]{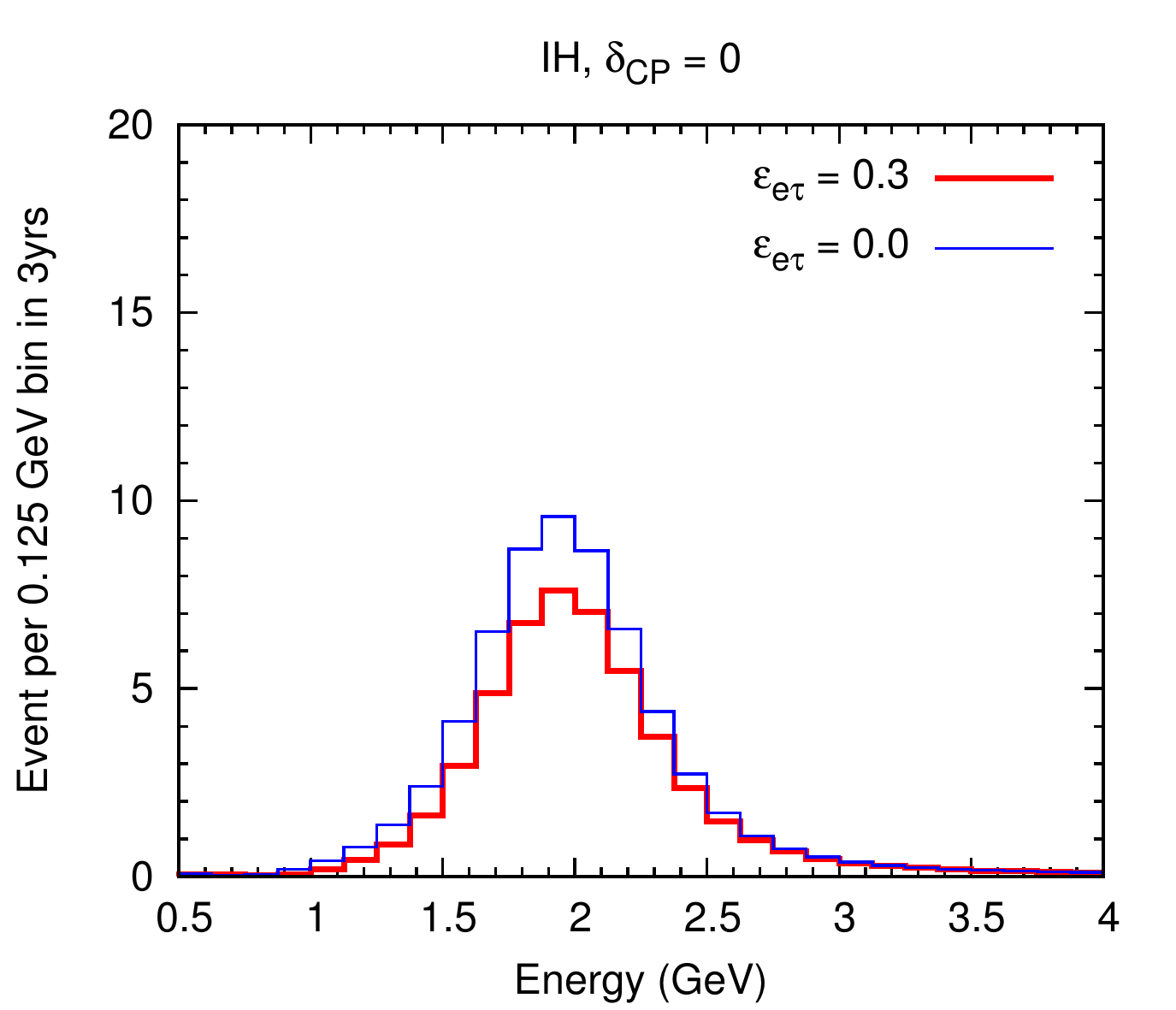}
\includegraphics[width=5.4cm,height=4.5cm]{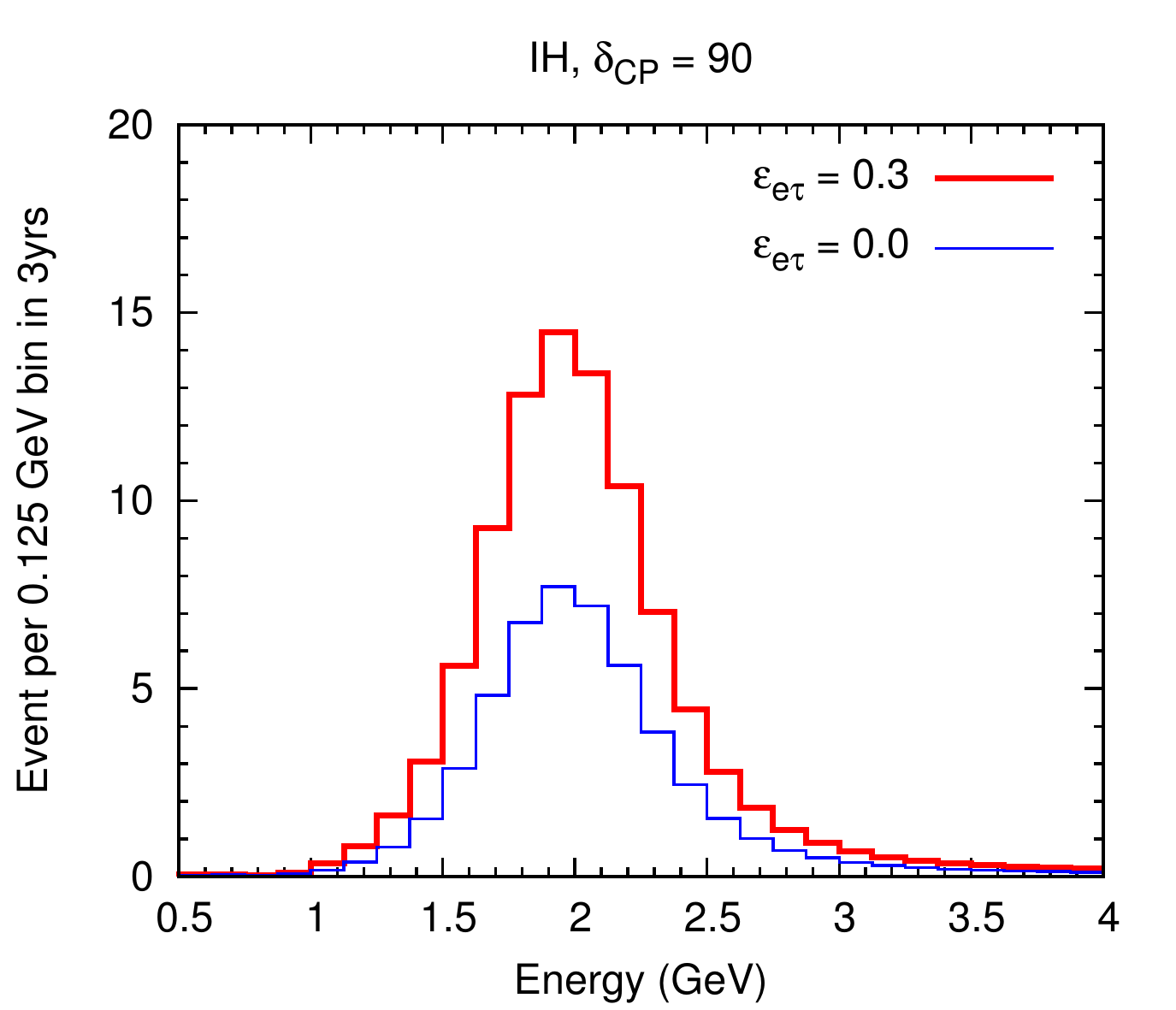}
\includegraphics[width=5.4cm,height=4.5cm]{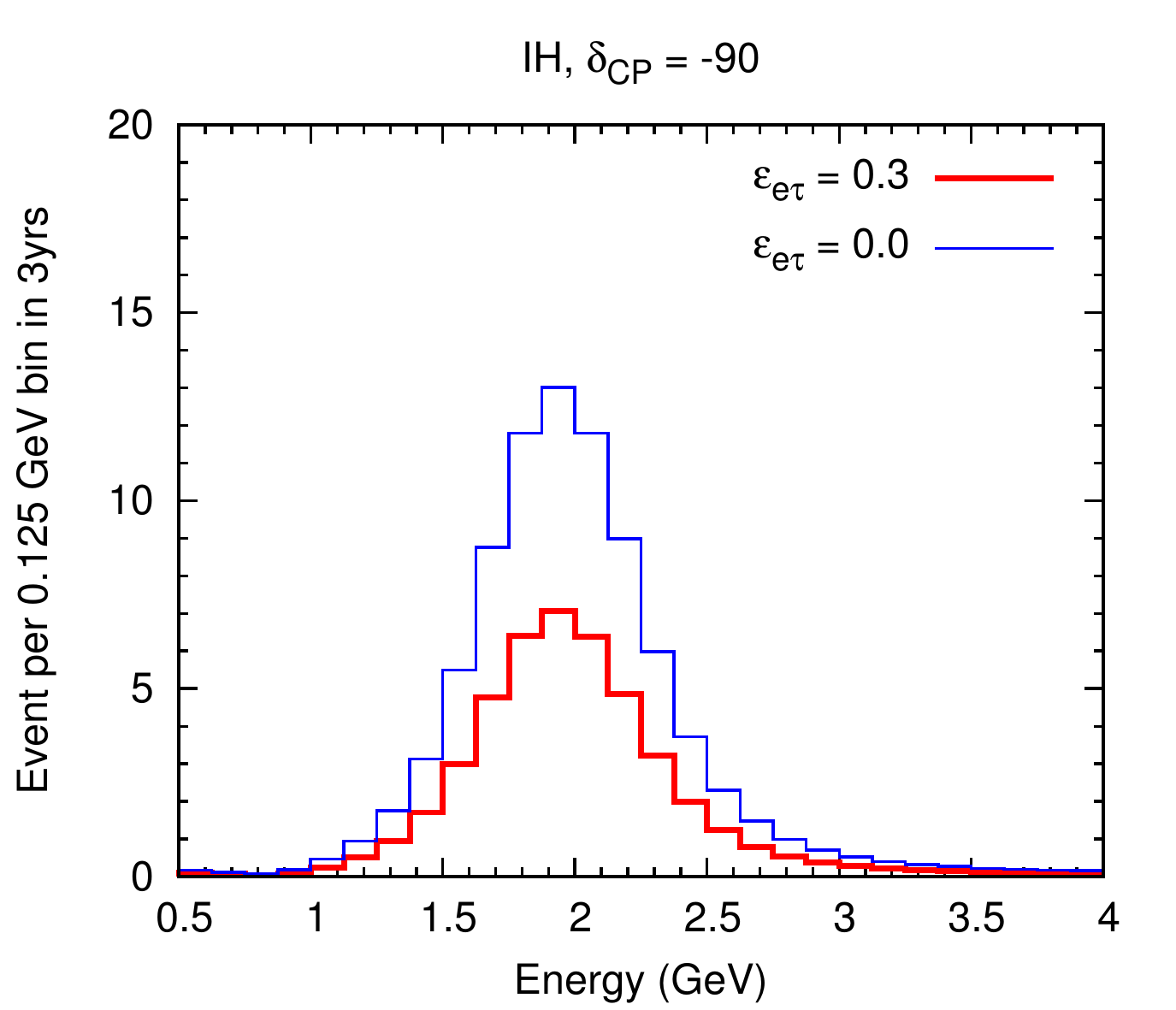}
\end{center}
\caption{The event spectra of NO$\nu$A for different values of CP violating phase, i.e, $\delta_{CP} = 0^{\circ}$ (left panel), $\delta_{CP} = 90^{\circ}$ (middle panel), and $\delta_{CP} = -90^{\circ}$ (right panel) .}
\label{novaspectra}
\end{figure}

\begin{figure}[!htb]
\begin{center}
\includegraphics[width=5.4cm,height=4.5cm]{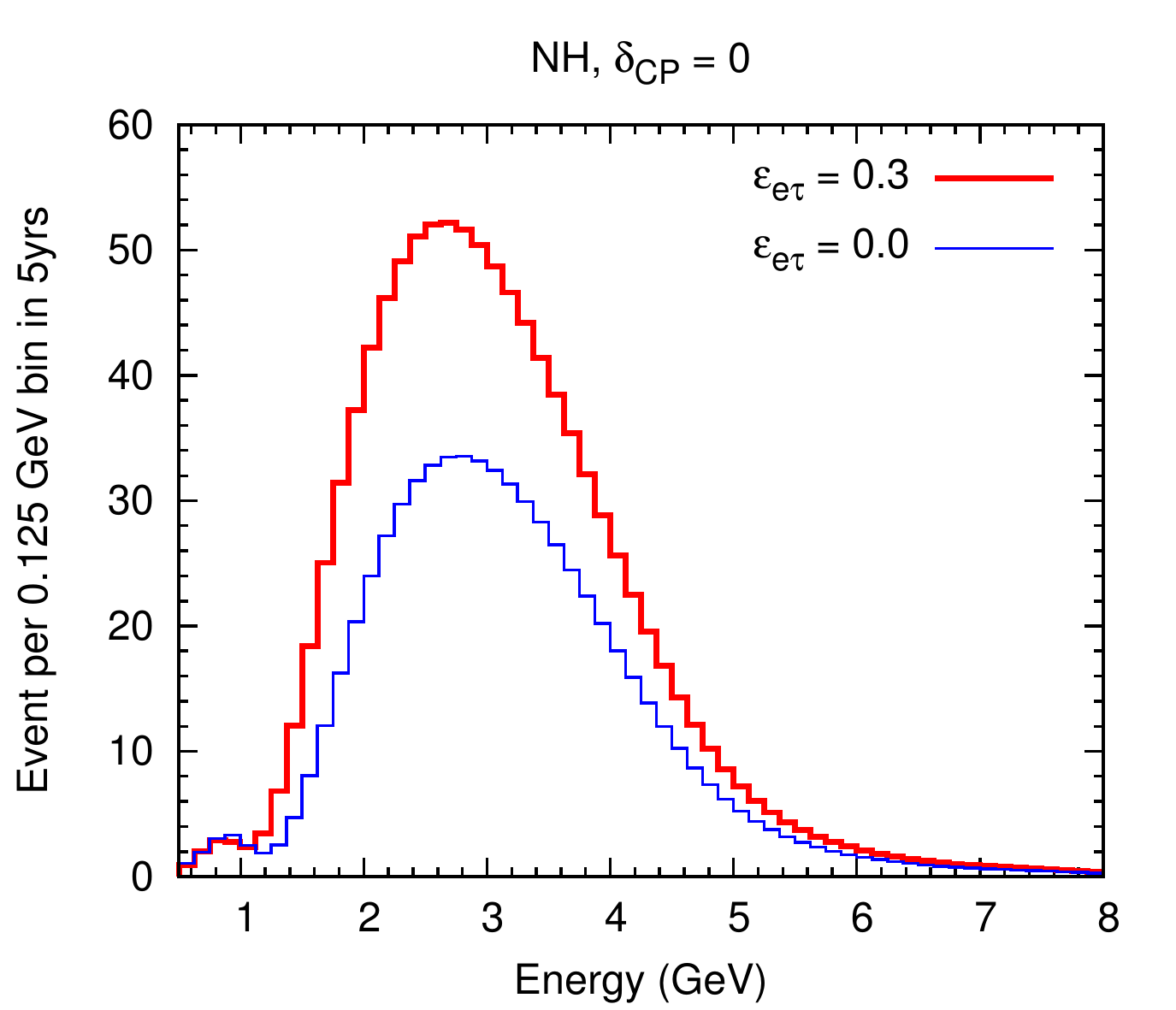}
\includegraphics[width=5.4cm,height=4.5cm]{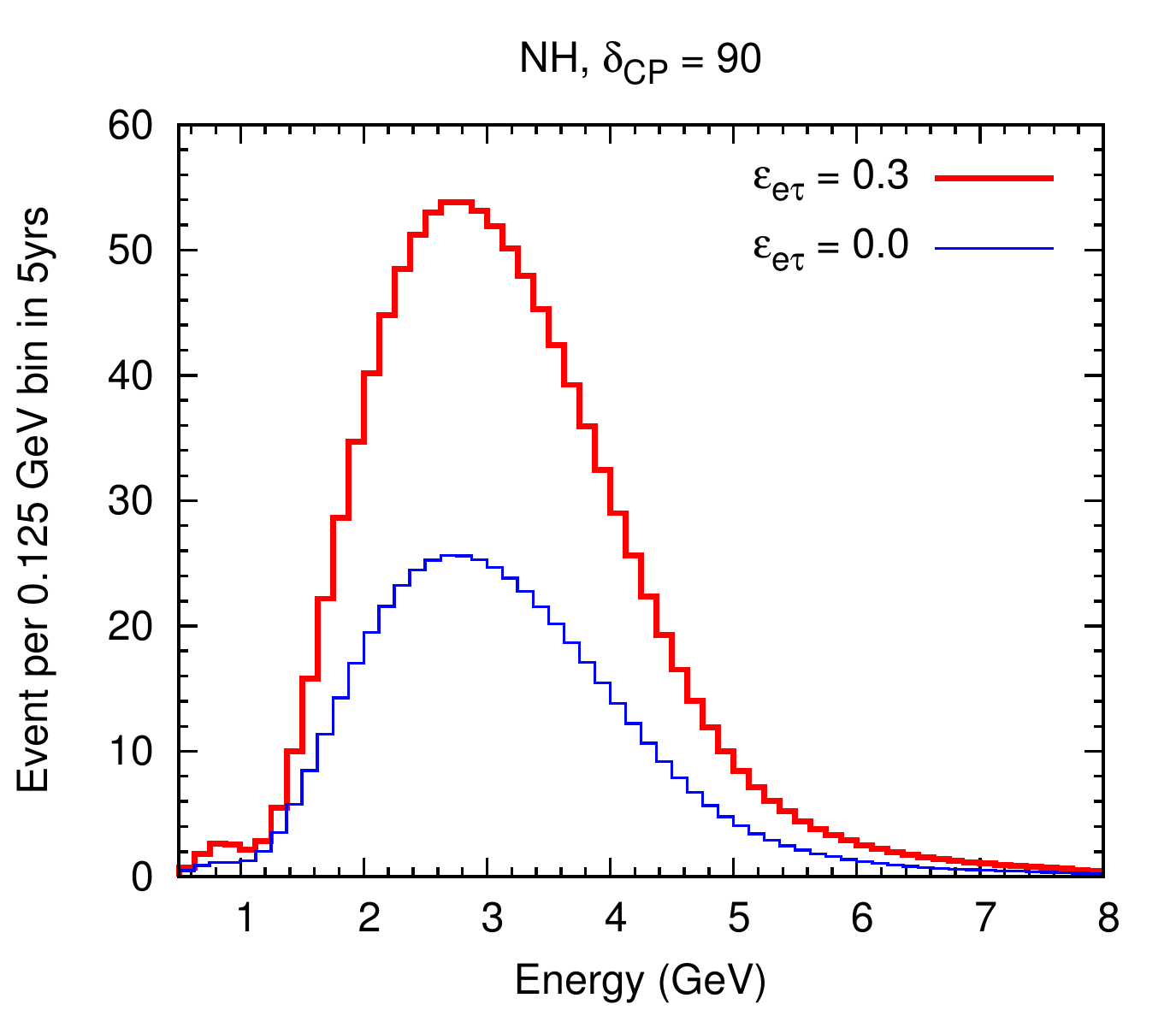}
\includegraphics[width=5.4cm,height=4.5cm]{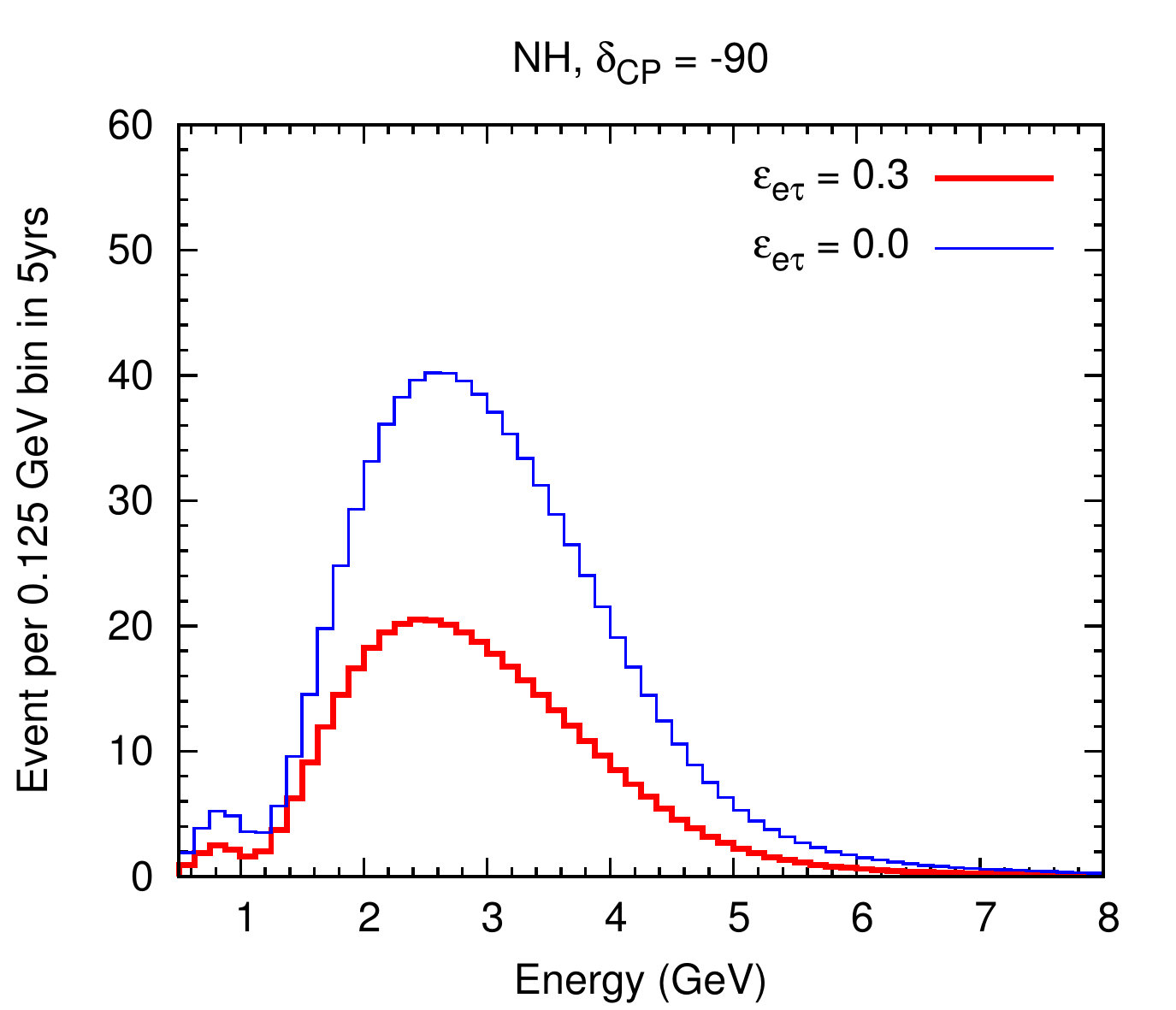}
\includegraphics[width=5.4cm,height=4.5cm]{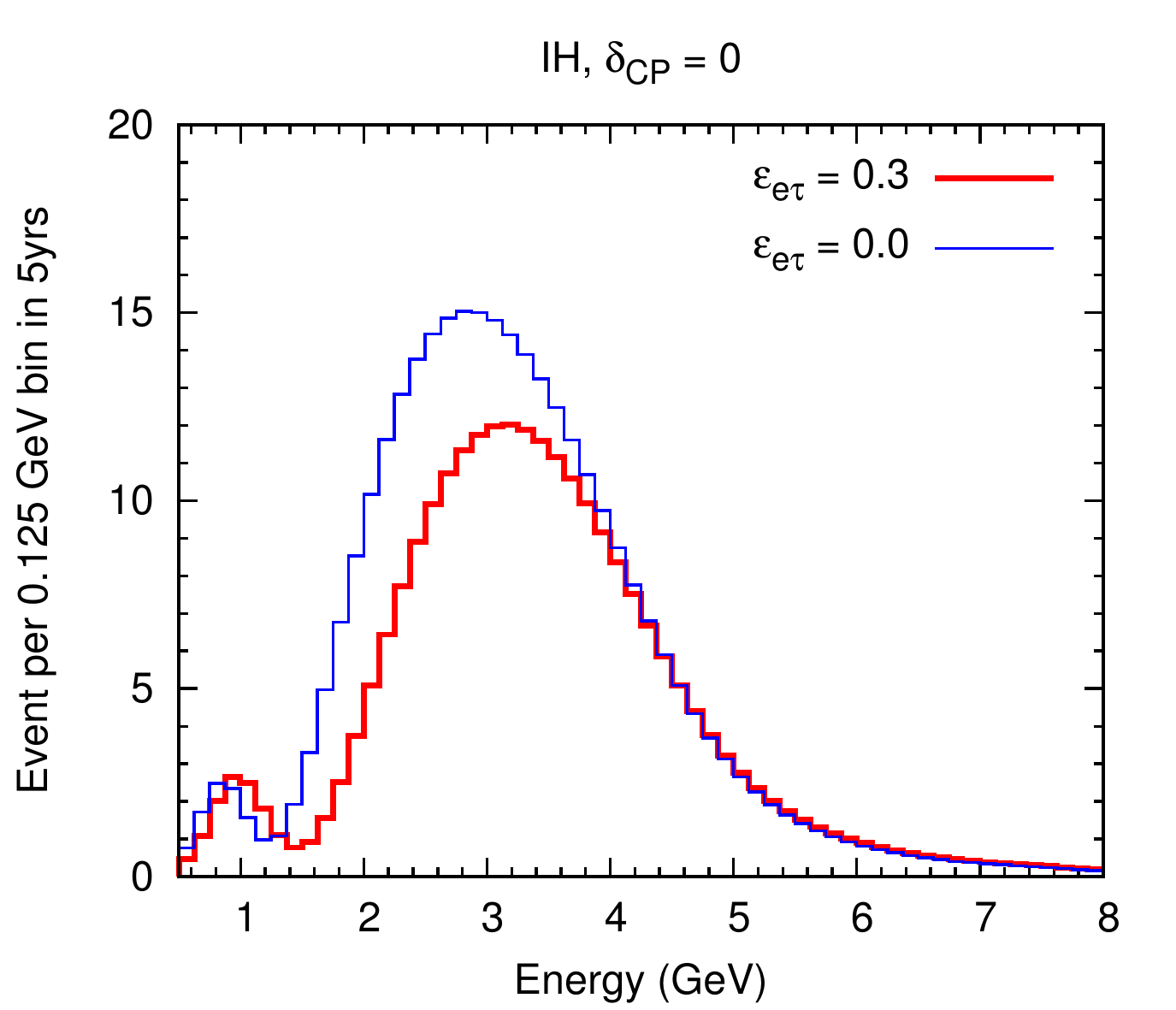}
\includegraphics[width=5.4cm,height=4.5cm]{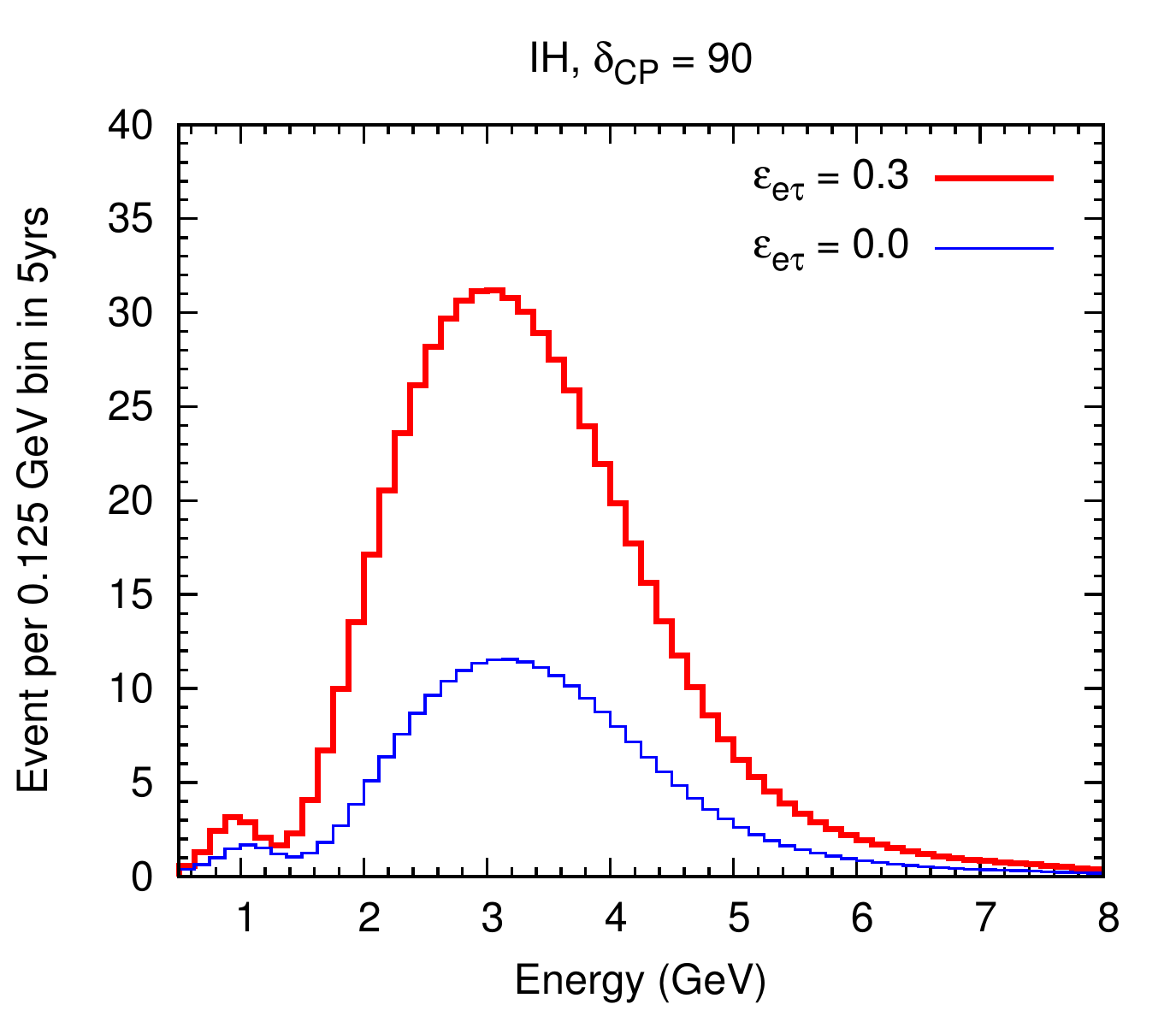}
\includegraphics[width=5.4cm,height=4.5cm]{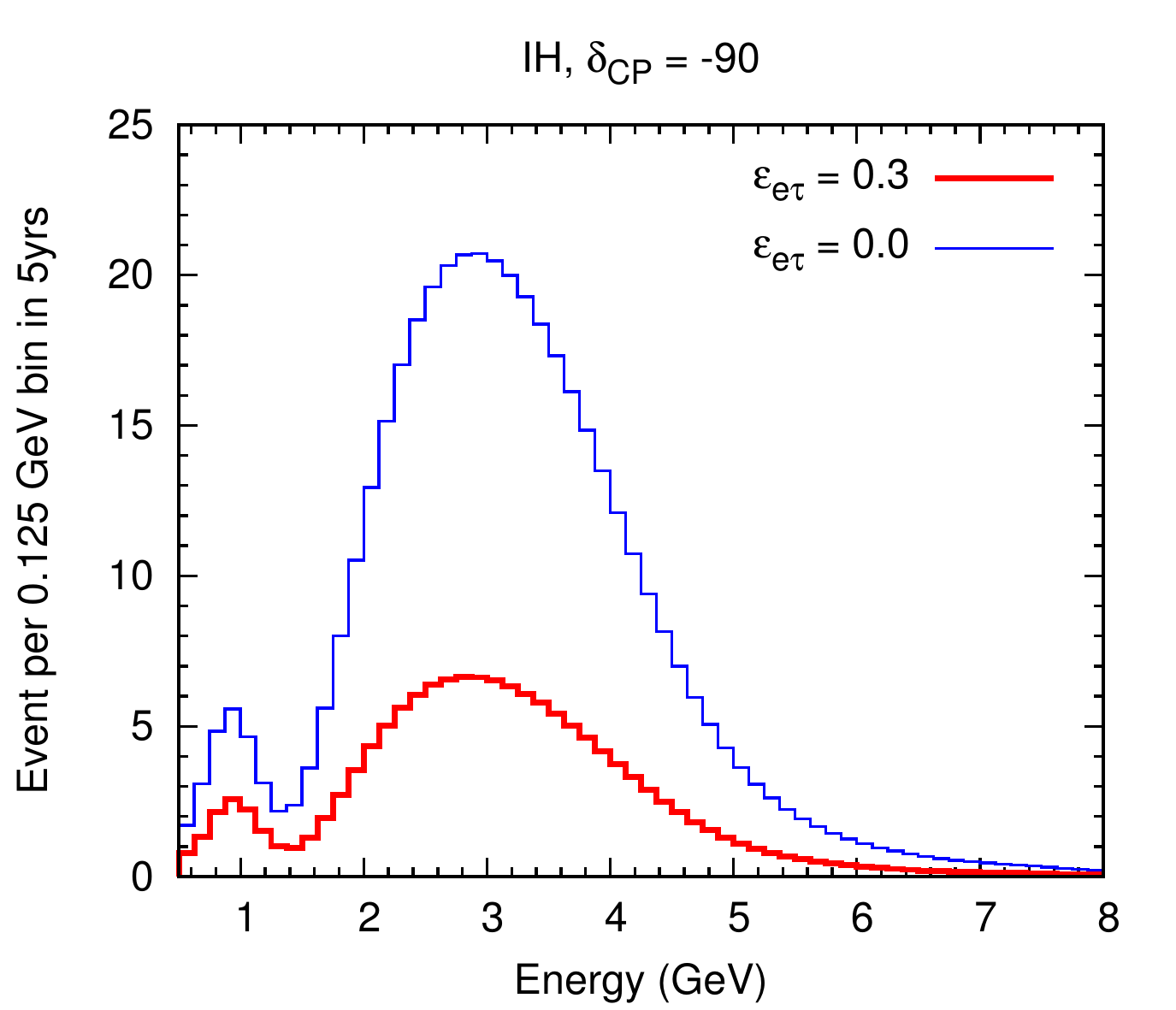}
\end{center}
\caption{The event spectra of DUNE for different values of CP violating phase, i.e, $\delta_{CP} = 0^{\circ}$ (left panel), $\delta_{CP} = 90^{\circ}$ (middle panel), and $\delta_{CP} = -90^{\circ}$ (right panel) .}
\label{dunespectra}
\end{figure}

Next, we show the oscillation probabilities  as a function of CP- violating phase for NO$\nu$A (DUNE) in the left (right)  panel of Fig.~\ref{app-disapp}. The dark solid (dashed) curve in the figure corresponds to oscillation probability for NH (IH) in the presence of NSI, whereas the light solid (dashed) curve corresponds to oscillation probability for NH (IH) in the standard oscillation. From the figure, we can see that there is an enhancement (diminution) in the probability for  CP- violating phase in the range $0^{\circ} \leqslant \delta _{CP} \leqslant 180^{\circ}$ ($ -180^{\circ} \leqslant \delta _{CP} \leqslant0^{\circ}$) for both mass hierarchies, if the NSI phase $\delta _{e\tau}$ is zero.
Further, the $\nu_e$ appearance event spectra for NO$\nu$A and DUNE  are shown in  Figs. \ref{novaspectra} and \ref{dunespectra} respectively. From these figures, we can see that the event rate in the presence of NSI  is larger than that in SO for $\delta_{CP}=$0 or $90^\circ$. Whereas for $\delta_{CP}=-90^\circ$, the event rates in presence of NSI  is lesser than that in SO for $\delta_{e\tau} =0$.
    
\subsection{Effect of NSI parameter on $\delta_{CP}$ sensitivity}
\begin{figure}[!htb]
\begin{center}
\includegraphics[width=8cm,height=6.5cm]{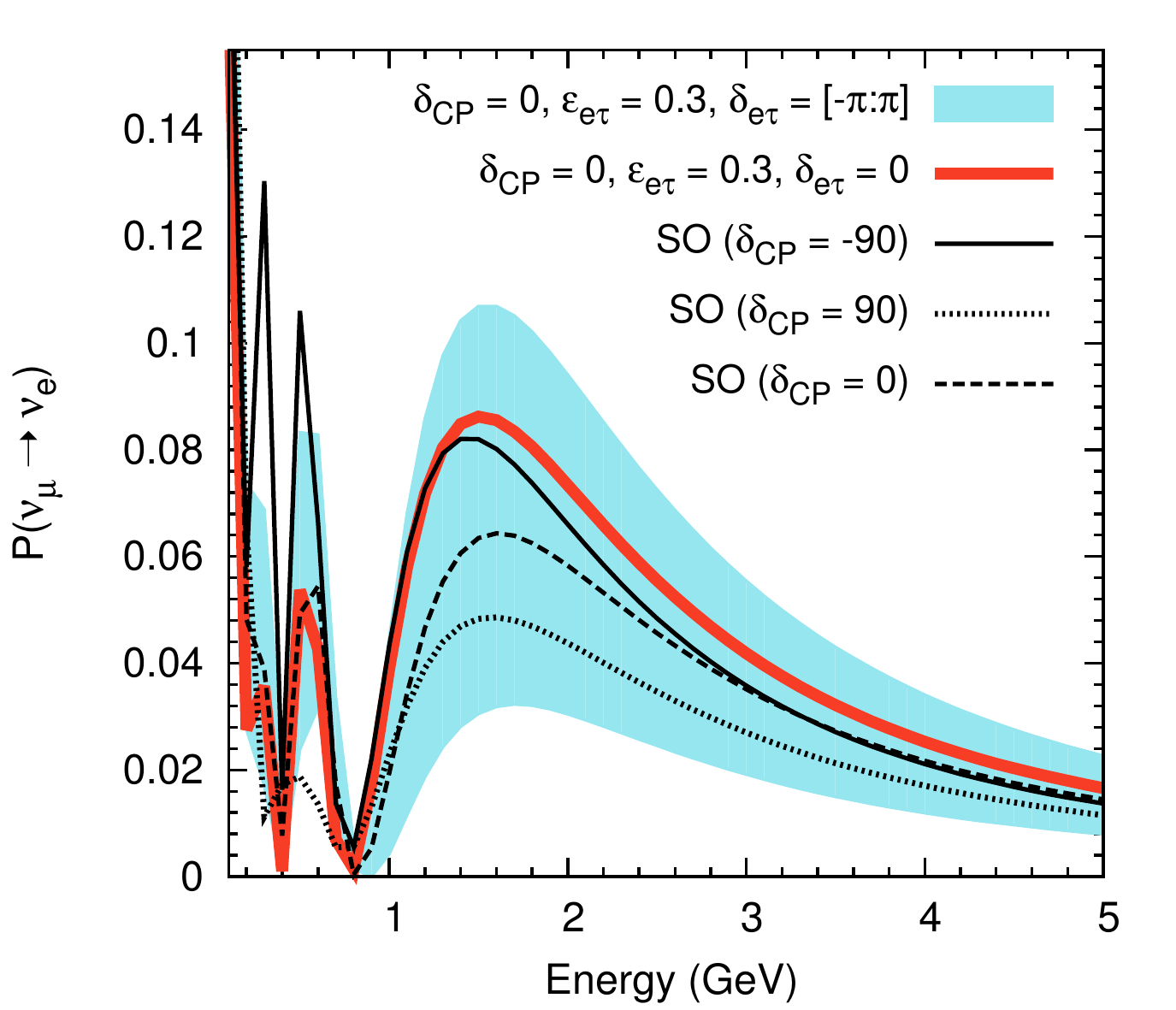}
\hspace*{0.1 truecm}
\includegraphics[width=8cm,height=6.5cm]{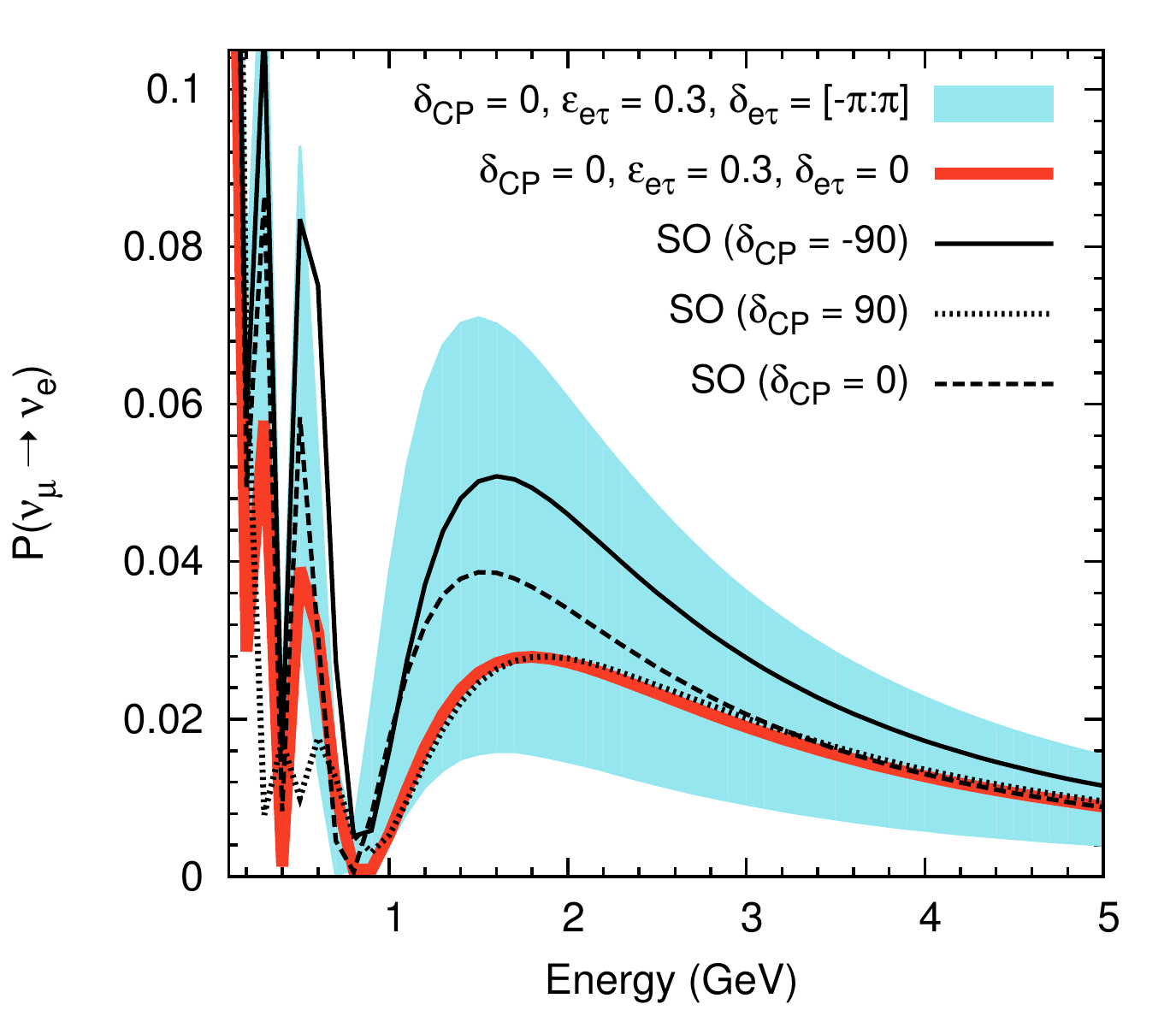}\\
\includegraphics[width=8cm,height=6.5cm]{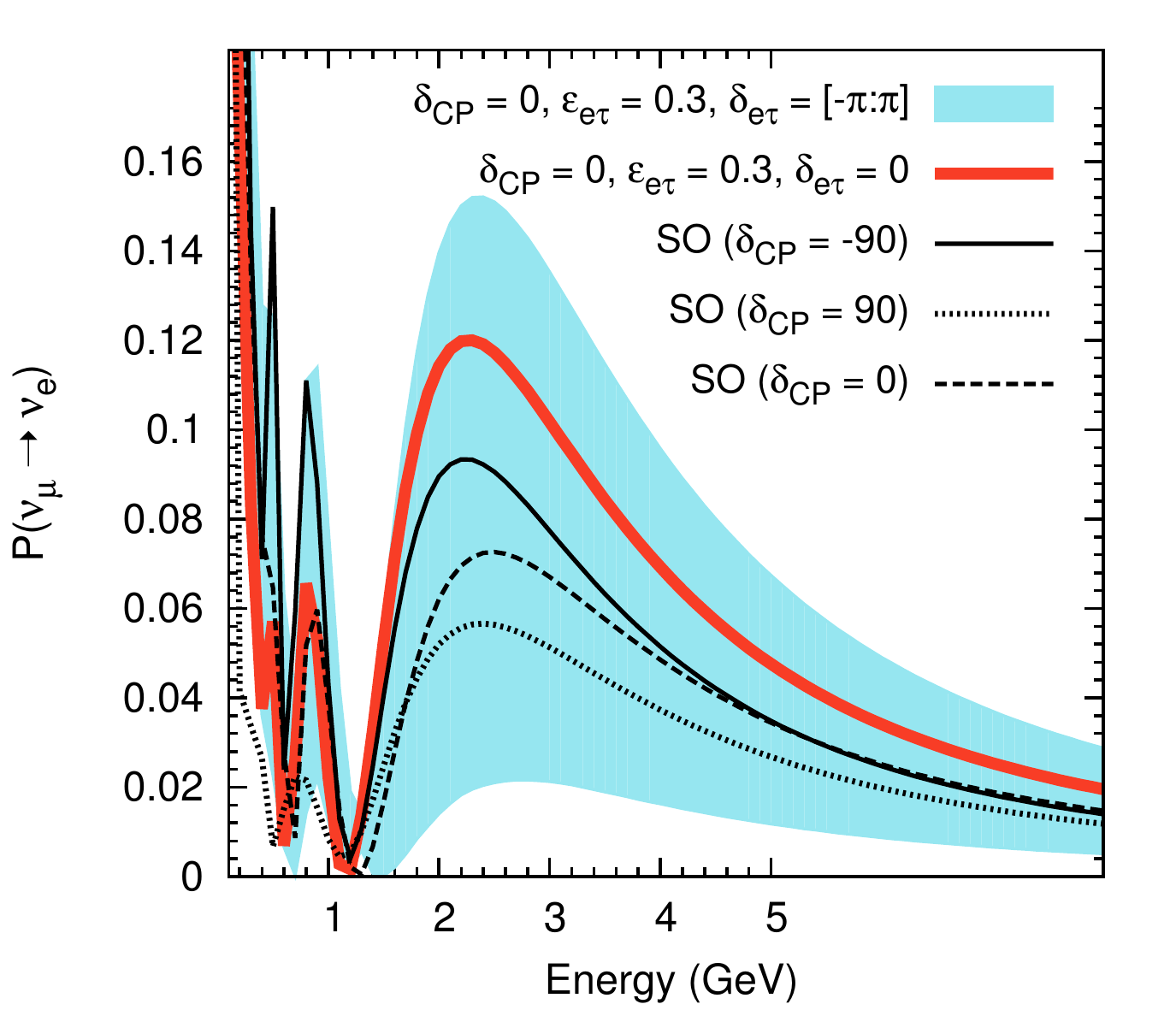}
\hspace*{0.1 truecm}
\includegraphics[width=8cm,height=6.5cm]{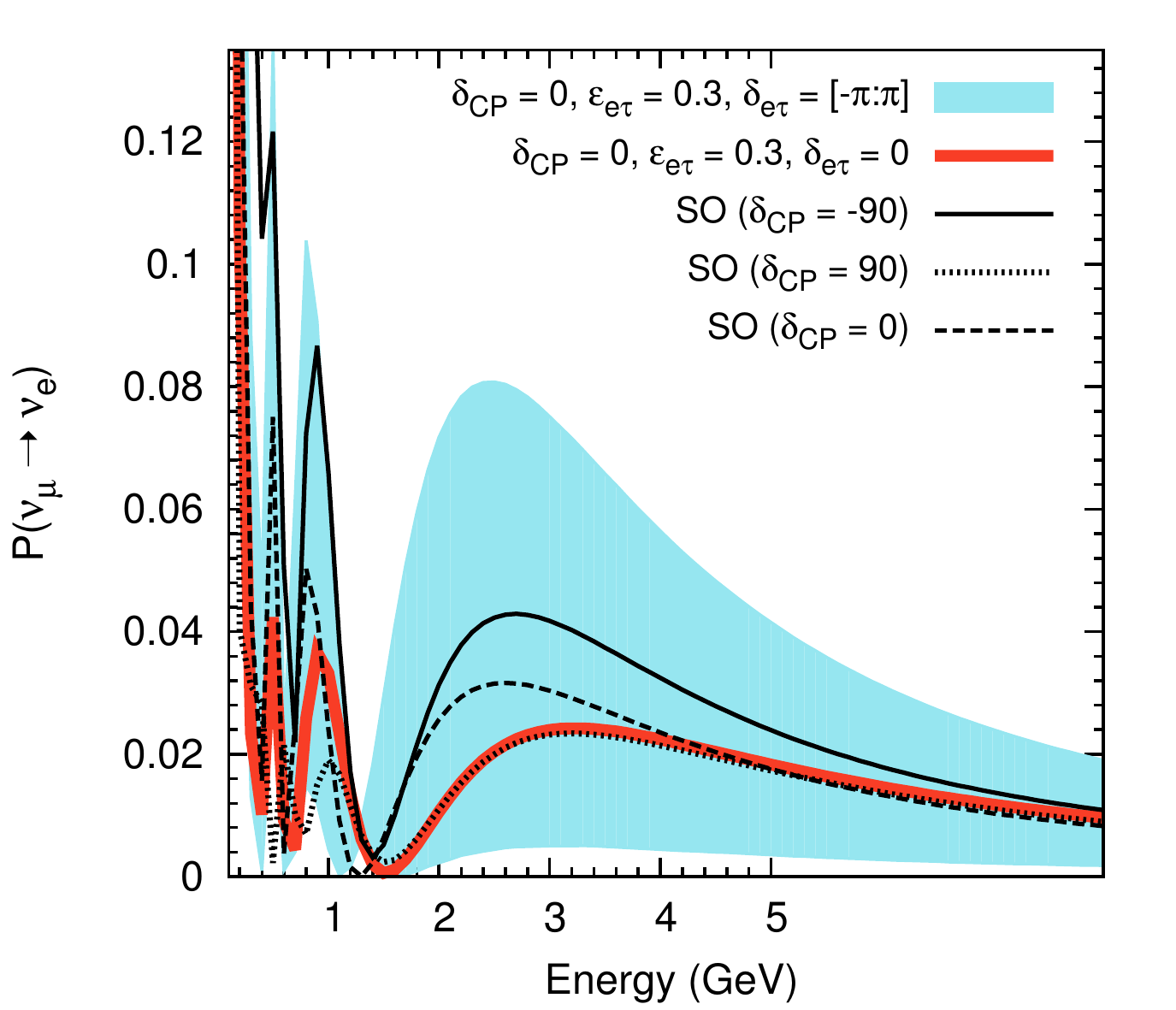}
\end{center}
\caption{ The $\nu_{\mu} \rightarrow \nu_e$  oscillation probability as a function of neutrino energy for NO$\nu$A (DUNE) in the top (bottom) panel.The left (right) panel corresponds to NH (IH).}
\label{osc}
\end{figure}
\begin{figure}[!htb]
\begin{center}
\includegraphics[width=7.5cm,height=6cm]{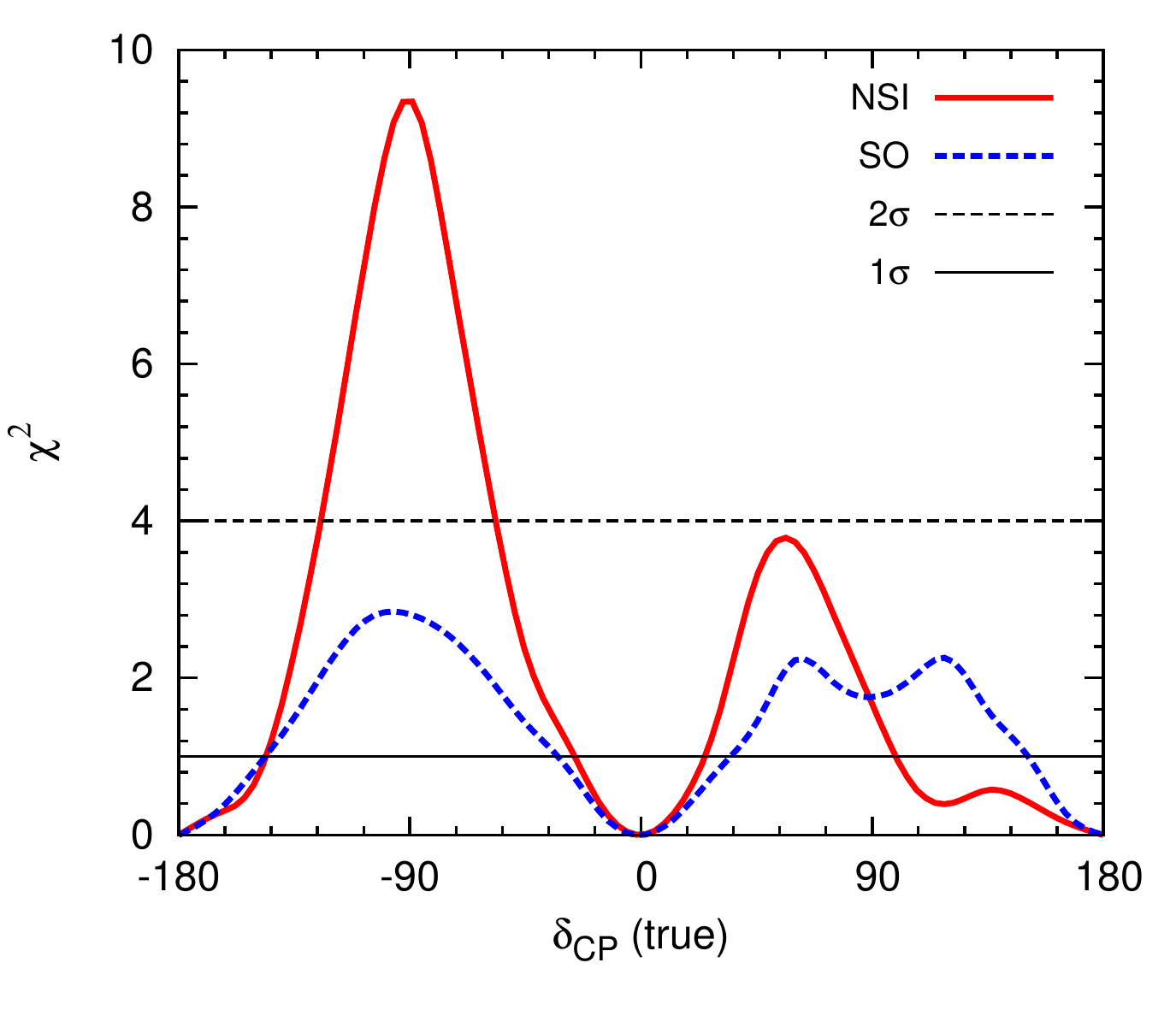}
\hspace*{0.15 truecm}
\includegraphics[width=7.5cm,height=6cm]{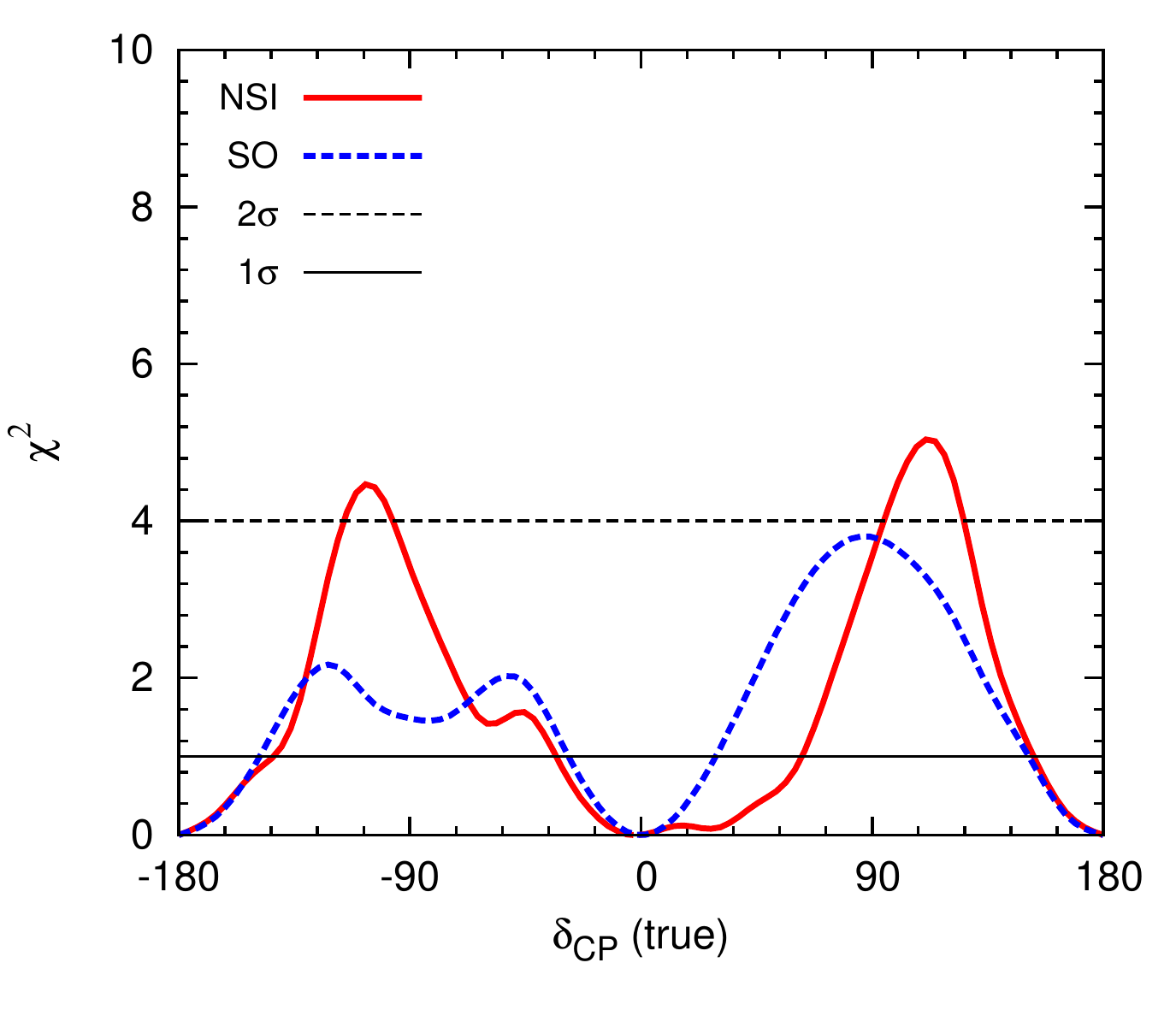}\\
\includegraphics[width=7.5cm,height=6cm]{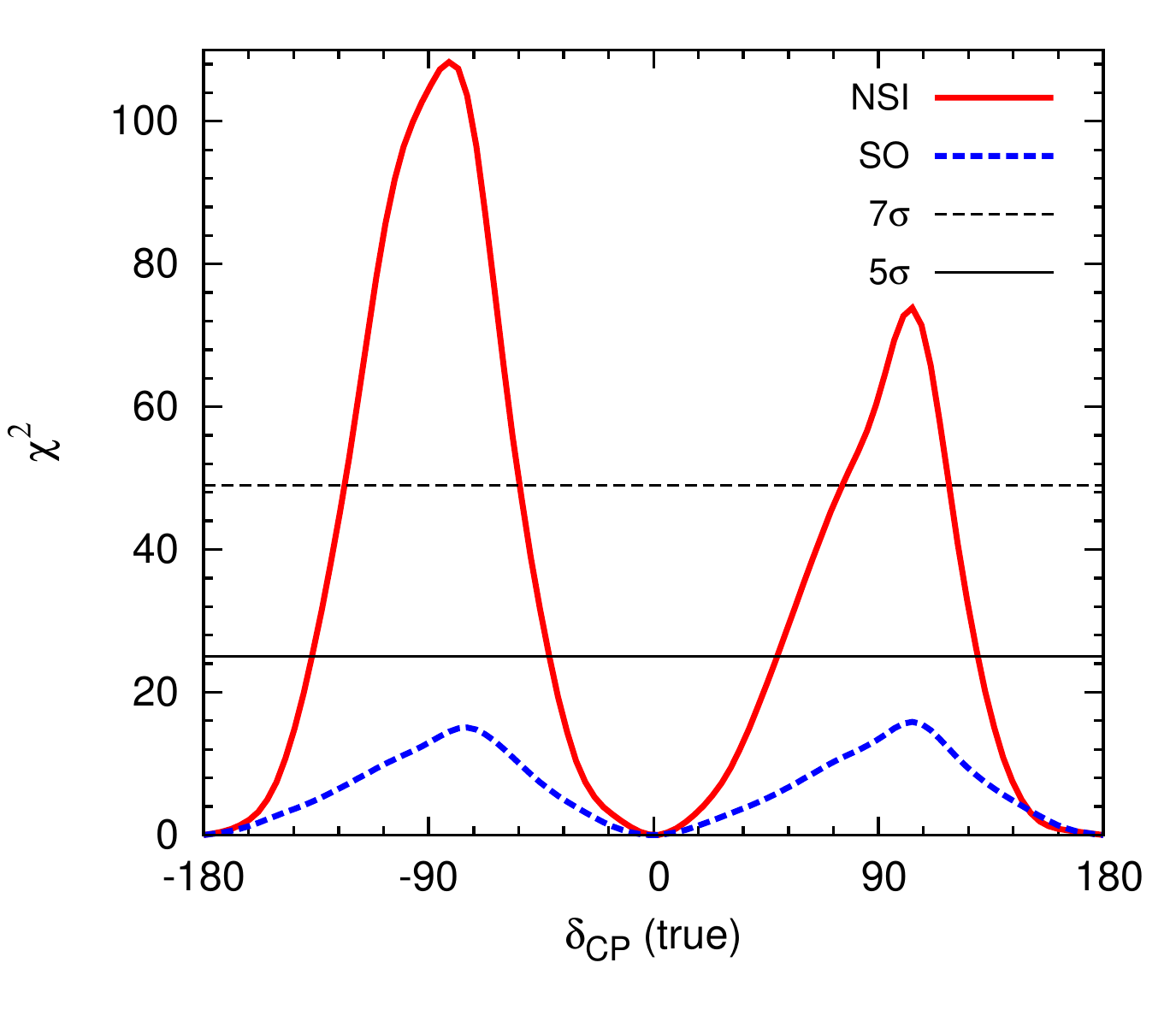}
\hspace*{0.15 truecm}
\includegraphics[width=7.5cm,height=6cm]{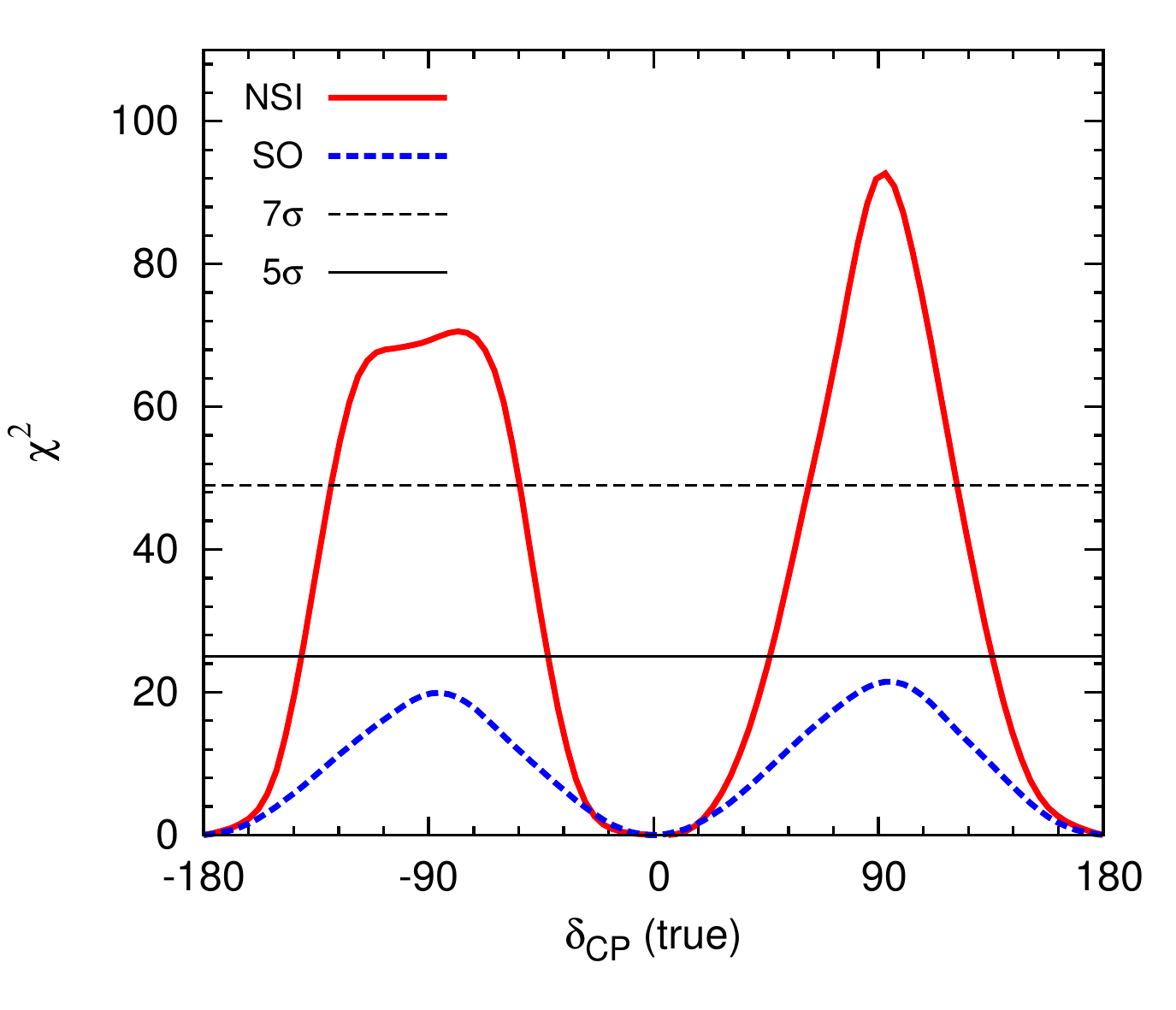}
\end{center}
\caption{ CP-violation sensitivity as a function of $\delta_{CP}$ for NO$\nu$A (DUNE) in the top (bottom) panel.The left (right) panel corresponds to NH (IH).}
\label{CPV}
\end{figure}

The neutrino oscillation probability for the channel $\nu_{\mu}\rightarrow\nu_e$ for NO$\nu$A  (DUNE) is given in the top (bottom) panels of Fig. \ref{osc}. The light coloured band in  the figure corresponds to the oscillation probability in the presence of NSI for allowed  values of NSI phase parameter $\delta_{e\tau}$  if $\delta_{CP}=0$. From the figure, we can see that the CP-violating oscillation signals (dark solid and dashed oscillation curves) in SO can mimic the CP-conserving oscillation signal (light solid oscillation curve) in presence of NSI. This leads to misinterpretation of oscillation data if NSIs exists in nature.  The CP-violation sensitivity ($\chi^2 = \chi^2 ({\delta_{CP}^{true}}) -\chi^2({\delta_{CP}^{test}=0,180})$) as a function of $\delta_{CP}$ for NO$\nu$A (DUNE) is given in the top (bottom) panel of Fig. \ref{CPV}. The dark solid curve in the figure corresponds to CPV sensitivity in presence of NSI, whereas the dark dashed curve in the figure corresponds to CPV sensitivity in SO. From the figure, we can see that NSI can significantly affect CPV sensitivity of both experiments. Though there is significant enhancement in the CPV sensitivity in presence of NSI for NO$\nu$A, it should be noted that the $\delta_{CP}$ coverage for CPV sensitivity above 1$\sigma$ is reduced in presence of  NSI while comparing with that of SO. Whereas for DUNE, the CPV sensitivity is enhanced in the presence of NSI and it is above 5$\sigma$ for more than 50\% allowed values of $\delta_{CP}$ in the case of both  NH and IH.
\section{Summary and Conclusions}
Conservation of lepton flavour universality is one of the unique feature of the SM.
However, recently there are a series of  experimental results in $B$ physics pointing towards possible
violations of LFU, both in the charged and neutral current mediated semileptonic decays. Such lepton flavour universality 
violation could in principle also
induce lepton flavour violating interactions. Considering the lepton flavour violating decays of $B$ meson, i.e, 
$B_d \to \tau^\pm e^\mp$ decay, we constrain the lepton flavour violating couplings in the $Z'$ model using
the upper limits of the corresponding branching ratios. We  obtained the bound  $|\varepsilon_{e \tau}| < 0.7$
from the decay rate. Assuming these NSI parameters in the charged lepton sectors to be related to the corresponding NSI parameters in the
neutrino sector by $SU(2)_L$ symmetry, we have studied the possible implications  of these new physics interactions in the long-baseline
neutrino oscillation experiments. In our analysis considering a conservative representative value for $\varepsilon_{e \tau}$ as
$\varepsilon_{e \tau}=0.3$ and we have investigated its implications in the  CP-violation sensitivity of long-baseline experiments. We found that the NSI parameters in the $e \tau$ sector remarkably affect the $\nu_e$ appearance oscillation probability. Moreover,  we found that the presence of NSIs lead to misinterpretation of oscillation data. The $\delta_{CP}$ coverage of NO$\nu$A for CPV sensitivity  above 1$\sigma$ is reduced in presence NSIs. However, the CPV sensitivity is enhanced in the presence of NSI and it is above 5$\sigma$ for more than 50\% allowed values of $\delta_{CP}$ in the case of both  NH and IH for DUNE.   
   
{\bf Acknowledgments}
We would like to thank Science and Engineering Research Board (SERB),
Government of India for financial support through grant No. SB/S2/HEP-017/2013. SC would like to thank Dr. Sushant K Raut, Dr. Arnab Dasgupta, Dr. Monojit Gosh and Mr. Mehedi Masud for many useful discussions regarding GLoBES.

\end{document}